\documentclass[%
 reprint,
 superscriptaddress,
 amsmath,amssymb,
 aps,
 prl,
prstper,
floatfix,
longbibliography 
]{revtex4-2}

\usepackage[pdftex]{graphicx} \graphicspath{{}}
\usepackage{tikz}
\usepackage{physics}
\usetikzlibrary{decorations.pathreplacing}
\usetikzlibrary{decorations.markings}
\usetikzlibrary{shapes.geometric}
\usepackage{subfigure}
\usepackage{float}
\usepackage{dcolumn}
\usepackage{bm}
\usepackage[pdftex,colorlinks=true]{hyperref}
\hypersetup{
  colorlinks=true,
  linkcolor=blue,
  urlcolor=cyan,
}

\tikzset{
  forwardarrow/.style={
    postaction={decorate},
    decoration={
      markings,
      mark=at position 0.5 with {\arrow{>}}
    }
  },
  backarrow/.style={
    postaction={decorate},
    decoration={
      markings,
      mark=at position 0.5 with {\arrow{<}}
    }
  },
  doublearrow/.style={
    postaction={decorate},
    decoration={
      markings,
      mark=at position 0.43 with {\arrow{<}},
      mark=at position 0.57 with {\arrow{>}}
    }
  }
}

\newcommand{\eqn}[1]{\begin{equation} #1 \end{equation}}
\newcommand{\eqa}[1]{\begin{align} #1 \end{align}}
\def\be{\begin{equation}}
\def\ee{\end{equation}}

\newcommand{\tik}[1]{\begin{tikzpicture} #1 \end{tikzpicture}}

\usepackage{color}
\definecolor{ForestGreen}{RGB}{34, 139, 34}

\newcommand{\nn}{\nonumber}

\newcommand{\mZ}{\mathcal{Z}}

\newcommand{\mF}{\mathcal{F}}
\newcommand{\mB}{\mathcal{B}}
\newcommand{\avg}[1]{\left\langle #1 \right\rangle}
\newcommand{\pd}{\partial}

\newcommand{\br}{\boldsymbol{r}}

\newcommand{\bk}{\boldsymbol{k}}

\newcommand{\bq}{\boldsymbol{q}}

\newcommand{\brho}{\boldsymbol{\rho}}
\newcommand{\bx}{\boldsymbol{\hat{x}}}
\newcommand{\by}{\boldsymbol{\hat{y}}}

\newcommand{\mO}{\mathcal{O}}

\newcommand{\mA}{\mathcal{A}}
\newcommand{\mG}{\mathcal{G}}

\newcommand{\hS}{\hat{S}}

\newcommand{\mL}{\mathcal{L}}

\newcommand{\hbS}{\hat{\boldsymbol{S}}}

\newcommand{\hH}{\hat{H}}

\newcommand{\barphi}{\bar{\phi}}

\newcommand{\had}{\hat{a}^{\dagger}}
\newcommand{\ha}{\hat{a}}

\newcommand{\sectionn}[1]{\textit{#1---}}

\begin{document}

\preprint{APS/123-QED}

%
\title{Exact Results on the Hydrodynamics of certain\\ Kinetically-Constrained Hopping Processes}

\author{Adam J. McRoberts}
\affiliation{International Centre for Theoretical Physics, Strada Costiera 11, 34151, Trieste, Italy}

\author{Vadim Oganesyan}
\affiliation{Physics Program and Initiative for the Theoretical Sciences, The Graduate Center, CUNY, New York, New York 10016, USA}
\affiliation{Department of Physics and Astronomy, College of Staten Island, CUNY, Staten Island, New York 10314, USA}

\author{Antonello Scardicchio}
\affiliation{International Centre for Theoretical Physics, Strada Costiera 11, 34151, Trieste, Italy}
\affiliation{INFN, Sezione di Trieste, Via Valerio 2, 34127 Trieste, Italy}

\date{\today}

\begin{abstract}
\noindent
We consider a model of interacting random walkers on a triangular chain and triangular lattice, where a particle can move only if the other two sites of the triangle are unoccupied -- a kinetically-constrained hopping process (KCHP) recently introduced in the context of non-linear diffusion cascades~\cite{raj2024diffusion, raj2024kinetically}. Using a classical-to-quantum mapping -- where the rate matrix of the stochastic KCHP corresponds to a spin Hamiltonian~\cite{raj2024kinetically}, and the equilibrium probability distribution to the quantum ground state -- we develop a systematic perturbation theory to calculate the diffusion constant; the hydrodynamics of the KCHPs is determined by the low-energy properties of the spin Hamiltonian, which we analyse with the standard Holstein-Primakoff spin-wave expansion.
For the triangular hopping we consider, we show that \textit{non-interacting} spin-wave theory predicts the \textit{exact} diffusion constant. 
We conjecture this holds for all KCHPs with (i) hard-core occupancy, (ii) parity-symmetry, and (iii) where the hopping processes are given by three-site gates -- that is, where hopping between two sites is conditioned on the occupancy of a third. 
We further show that there are corrections to the diffusion constant when the KCHP is described by \textit{four}-site gates, which we calculate at leading order in the semi-classical $1/S$ expansion. We support all these conclusions with numerical simulations.
\end{abstract}
\maketitle
The study of many-body systems that evince anomalously slow dynamics has long been a source of unexpected, and often subtle, phenomena. A particularly prominent example is the study of spin glasses~\cite{edwards1975theory,binder1986spin}, from which, through Parisi's theory of replica symmetry breaking~\cite{mezard1987spin}, arose a new paradigm in the study of complex systems~\cite{parisi2023nobel}. A similarly important example is the theory of Anderson localisation~\cite{anderson1958absence}, and its quantum many-body extension since the works of Basko, Aleiner, and Altshuler~\cite{basko2006metal,abanin2019colloquium,sierant2024many}. Common to both, however (aside from originating with Anderson), is that, despite their well-known phenomenology, analytical -- or conclusive numerical -- results are rare.

The closely related (and, arguably, harder) glass problem, in particular, has been the focus of intense study across several fields of chemistry, physics, and mathematics, ranging from large scale \textit{ab initio} studies of molecular dynamics~\cite{berthier2023modern}, to mode-coupling theory~\cite{das2004mode}, to statistical mechanics models of geometric frustration~\cite{sadoc1999geometrical,tarjus2005frustration}. Kinetically-constrained hopping processes (KCHPs) were introduced~\cite{ritort2003,biroli2013perspective} as models where glassiness arises purely from dynamics, i.e., where the equilibrium state is only weakly-correlated. Although some analytical techniques, and a few exact results (e.g. in 1D integrable models~\cite{felderhof1971spin}), are known, the principal advantage of KCHPs is the simplicity that facilitates large-scale numerical simulations. Somewhat counter-intuitively, perhaps, transport is not believed to be essential to understanding the glass problem, with progress in recent years focusing on understanding the role of jamming~\cite{mari2009jamming} and other examples of falling out of thermal equilibrium~\cite{garrahan2007dynamical}.

\begin{figure}[thb]
    \centering
    \tik{
    \draw(-0.2, 0.566) node[left]{$(a)$};
    \fill[gray] (0.0, 0.0) circle [radius=3pt];
    \draw[black] (0.0, 0.0) circle [radius=3pt];
    \draw[black] (0.4, 0.566) circle [radius=3pt];
    \fill[gray] (0.8, 0.0) circle [radius=3pt];
    \draw[black] (0.8, 0.0) circle [radius=3pt];
    \draw[->] (0.8 - 0.05, 0.1) arc (0:68:0.4);
    \draw[->] (0.8 + 0.05, 0.1) arc (180:112:0.4);
    \draw[->] (0.8 + 0.10, 0.1 - 0.04) arc (-72:0:0.4);
    \draw[->] (0.8 + 0.10, 0.0) arc (112:68:0.74);
    \draw[black] (1.2, 0.566) circle [radius=3pt];
    \draw[black] (1.6, 0.0) circle [radius=3pt];
    \fill[gray] (2.0, 0.566) circle [radius=3pt];
    \draw[black] (2.0, 0.566) circle [radius=3pt];
    \draw[->] (2.0 - 0.10, 0.566) arc (-68:-112:0.74);
    \draw[->] (2.0 - 0.05, 0.566 - 0.1) arc (0:-68:0.4);
    \draw[->] (2.0 + 0.05, 0.566 - 0.1) arc (-180:-112:0.4);
    \draw[->] (2.0 - 0.10, 0.566 - 0.1 + 0.04) arc (112:180:0.4);
    \draw[black] (2.4, 0.0) circle [radius=3pt];
    \fill[gray] (2.8, 0.566) circle [radius=3pt];
    \draw[black] (2.8, 0.566) circle [radius=3pt];
    \fill[gray] (3.2, 0.0) circle [radius=3pt];
    \draw[black] (3.2, 0.0) circle [radius=3pt];
    \fill[gray] (3.6, 0.566) circle [radius=3pt];
    \draw[black] (3.6, 0.566) circle [radius=3pt];
    \fill[gray] (4.0, 0.0) circle [radius=3pt];
    \draw[black] (4.0, 0.0) circle [radius=3pt];
    \draw[->] (4.0 + 0.10, 0.1 - 0.04) arc (-72:0:0.4);
    \draw[->] (4.0 + 0.10, 0.0) arc (112:68:0.74);
    \draw[black] (4.4, 0.566) circle [radius=3pt];
    \draw[black] (4.8, 0.0) circle [radius=3pt];
    \fill[gray] (5.2, 0.566) circle [radius=3pt];
    \draw[black] (5.2, 0.566) circle [radius=3pt];
    \draw[->] (5.2 - 0.10, 0.566) arc (-68:-112:0.74);
    \draw[->] (5.2 + 0.10, 0.566) arc (-112:-68:0.74);
    \draw[->] (5.2 - 0.05, 0.566 - 0.1) arc (0:-68:0.4);
    \draw[->] (5.2 + 0.05, 0.566 - 0.1) arc (-180:-112:0.4);
    \draw[->] (5.2 - 0.10, 0.566 - 0.1 + 0.04) arc (112:180:0.4);
    \draw[->] (5.2 + 0.10, 0.566 - 0.1 + 0.04) arc (68:0:0.4);
    \draw[black] (5.6, 0.0) circle [radius=3pt];
    \draw[black] (6.0, 0.566) circle [radius=3pt];
    %
    %
    \draw(-0.2, -1.5 + 0.566) node[left]{$(b)$};
    \fill[gray] (0.0, -1.5 + 0.0) circle [radius=3pt];
    \draw[black] (0.0, -1.5 + 0.0) circle [radius=3pt];
    \draw[black] (0.0, -1.5 - 2*0.566) circle [radius=3pt];
    \fill[gray] (0.4, -1.5 + 0.566) circle [radius=3pt];
    \draw[black] (0.4, -1.5 + 0.566) circle [radius=3pt];
    \fill[gray] (0.4, -1.5 - 0.566) circle [radius=3pt];
    \draw[black] (0.4, -1.5 - 0.566) circle [radius=3pt];
    \draw[->] (0.4 - 0.05, -1.5 - 0.566 - 0.1) arc (0:-68:0.4);
    \draw[->] (0.4 + 0.05, -1.5 - 0.566 - 0.1) arc (-180:-112:0.4);

    \draw[black] (0.8, -1.5 + 0.0) circle [radius=3pt];
    \draw[black] (0.8, -1.5 - 2*0.566) circle [radius=3pt];
    \fill[gray] (1.2, -1.5 + 0.566) circle [radius=3pt];
    \draw[black] (1.2, -1.5 + 0.566) circle [radius=3pt];
    \fill[gray] (1.2, -1.5 - 0.566) circle [radius=3pt];
    \draw[black] (1.2, -1.5 - 0.566) circle [radius=3pt];
    \draw[->] (1.2 - 0.05, -1.5 - 0.566 - 0.1) arc (0:-68:0.4);
    \draw[->] (1.2 + 0.05, -1.5 - 0.566 - 0.1) arc (-180:-112:0.4);
    \fill[gray] (1.6, -1.5 + 0.0) circle [radius=3pt];
    \draw[black] (1.6, -1.5 + 0.0) circle [radius=3pt];
    \draw[->] (1.6 + 0.10, -1.5 + 0.1 - 0.04) arc (-72:0:0.4);
    \draw[->] (1.6 + 0.10, -1.5 + 0.0) arc (112:68:0.74);
    \draw[black] (1.6, -1.5 - 2*0.566) circle [radius=3pt];
    \draw[black] (2.0, -1.5 + 0.566) circle [radius=3pt];
    \fill[gray] (2.0, -1.5 - 0.566) circle [radius=3pt];
    \draw[black] (2.0, -1.5 - 0.566) circle [radius=3pt];
    \draw[black] (2.4, -1.5 + 0.0) circle [radius=3pt];
    \fill[gray] (2.4, -1.5 - 2*0.566) circle [radius=3pt];
    \draw[black] (2.4, -1.5 - 2*0.566) circle [radius=3pt];
    \fill[gray] (2.8, -1.5 + 0.566) circle [radius=3pt];
    \draw[black] (2.8, -1.5 + 0.566) circle [radius=3pt];
    \draw[->] (2.8 - 0.05, -1.5 + 0.566 - 0.1) arc (0:-68:0.4);
    \draw[->] (2.8 + 0.05, -1.5 + 0.566 - 0.1) arc (-180:-112:0.4);
    \draw[->] (2.8 + 0.10, -1.5 + 0.566) arc (-112:-68:0.74);
    \draw[->] (2.8 + 0.10, -1.5 + 0.566 - 0.1 + 0.04) arc (68:0:0.4);
    \draw[->] (2.8 - 0.10, -1.5 + 0.566 - 0.1 + 0.04) arc (112:180:0.4);
    \draw[->] (2.8 - 0.10, -1.5 + 0.566) arc (-68:-112:0.74);
    \fill[gray] (2.8, -1.5 - 0.566) circle [radius=3pt];
    \draw[black] (2.8, -1.5 - 0.566) circle [radius=3pt];
    \draw[->] (2.8 - 0.05, -1.5 - 0.566 + 0.1) arc (0:68:0.4);
    \draw[->] (2.8 + 0.05, -1.5 - 0.566 + 0.1) arc (180:112:0.4);
    %
    \draw[black] (3.2, -1.5 + 0.0) circle [radius=3pt];
    \draw[black] (3.2, -1.5 - 2*0.566) circle [radius=3pt];
    \draw[black] (3.6, -1.5 + 0.566) circle [radius=3pt];
    \fill[gray] (3.6, -1.5 - 0.566) circle [radius=3pt];
    \draw[black] (3.6, -1.5 - 0.566) circle [radius=3pt];
    \draw[black] (5.0, -2.0) circle [radius=3pt];
    \draw(5.1, -2.0) node[right]{$n = 0 \sim \ket{\downarrow}$};
    \fill[gray] (5.0, -2.5) circle [radius=3pt];
    \draw[black] (5.0, -2.5) circle [radius=3pt];
    \draw(5.1, -2.5) node[right]{$n = 1 \sim \ket{\uparrow}$};
    }
    \caption{Sketch of the hopping rules for $(a)$ the triangular chain and $(b)$ triangular lattice kinetically-constrained hopping models. At each stage of the process, a random triangle is chosen, and if there is exactly one particle on the triangle it hops to one of the other two sites with equal probability. Possible hops are shown with arrows; two arrows are shown if it could proceed via one of two different triangles. This model is similar but not identical to the one studied in Refs.~\cite{jackle1994kinetic,kronig1994kinetic}.}
    \label{fig:triangle_hopping_rules}
\end{figure}

\begin{figure*}[thb]
    \centering
    \includegraphics[width=\linewidth]{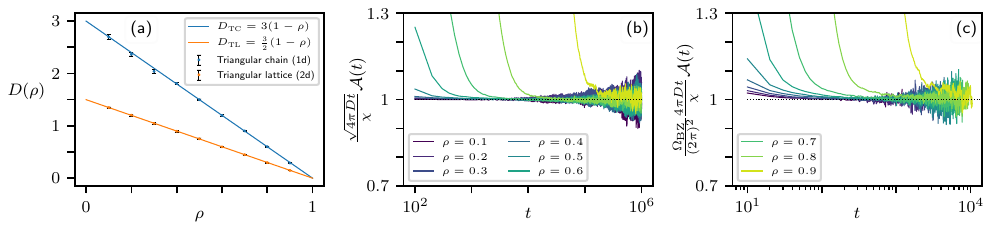}
    \caption{Lack of corrections to the bare diffusion constant, obtained from non-interacting spin wave theory, in the triangular chain and triangular lattice models. (a) Numerical estimates of the diffusion constants. (b), (c) Convergence of the autocorrelator to the non-interacting spin wave theory predictions, $D_{\mathrm{TC}} = 3(1 - \rho)$ in the triangular chain, and $D_{\mathrm{TL}} = \frac{3}{2}(1 - \rho)$ in the triangular lattice, respectively. $N = 8192$ for the chain; $N = 256^2$ for the triangular lattice, with PBCs and $10^4$ states in the ensemble in all cases. The finite time corrections also appear numerically to qualitatively agree with Ref.~\cite{michailidis2024corrections}.
    }
    \label{fig:triangular_diffusion}
\end{figure*}

Outwith the realm of the glass problem, however, the transport properties which arise from the constrained dynamics of KCHPs are interesting in their own right, and not always immediately apparent; we mention, in passing, that for the models of constrained-hopping on triangles shown in Fig.~\ref{fig:triangle_hopping_rules} there are exponentially many jammed configurations for $\rho > 2/3$, but this turns out not to affect the near-equilibrium transport~\cite{raj2024kinetically}.

In this Letter, we exploit the mapping of a classical, stochastic KCHP to an interacting ${S = 1/2}$ Hamiltonian and extract its hydrodynamics from the low-energy properties of the quantum theory -- which we analyse as a semi-classical series in powers of $1/S$ with Holstein-Primakoff spin-wave theory. 
We delineate a 
class of KCHPs for which the bare diffusion constant -- obtained from non-interacting spin-wave theory -- is \textit{exact}, including a triangular chain recently introduced in the context of non-linear diffusion cascades~\cite{raj2024diffusion, raj2024kinetically,delacretaz2020heavy}, and its extension to the triangular lattice (cf. Fig.~\ref{fig:triangle_hopping_rules}).

For other KCHPs, we show how corrections to the diffusion constant can be calculated using Feynman diagrams. This mapping and effective field theory are not limited by the lattice geometry or spatial dimensionality, and, indeed, the higher-order terms become more irrelevant (in the RG sense) as the dimension increases.  
Our work ties together prior effort to study transport in KCHPs~\cite{arita2018bulk,blondel2015tracer} and the classical-to-quantum mapping~\cite{peliti1985path}, which was previously utilized (in reverse!) in the context of quantum dimer models~\cite{henley1997relaxation}.

\sectionn{Triangular models \& classical-to-quantum mapping}
A classical stochastic process (such as the kinetically-constrained hopping we consider here) is described by a rate matrix $H$ which governs the evolution of the probability distribution through the master equation,
\eqn{
\frac{dP(\brho, t)}{dt} = -\sum_{\brho'}H_{\brho,\brho'}P(\brho', t),
}
where $P(\brho, t)$ is the probability that the system has the configuration $\brho = \{\rho_1,\,...\,,\rho_N\}$ at time $t$, with $\rho_j = 1$ and $\rho_j = 0$ denoting a particle and hole, respectively.

Assuming the dynamics is diffusive, the diffusion constant $D$ is defined by the autocorrelator,
\eqa{
\mA(t) &:= \frac{1}{N}\sum_j\avg{(\rho_j(t) - \rho)(\rho_j(0) - \rho)} \nn
\\
&= \frac{(2\pi)^d}{\Omega_{\mathrm{BZ}}}\frac{\chi}{\left(4\pi D t\right)^{d/2}} +\ldots\,,
\label{eq:autocorrelator}
}
where $\rho = \avg{\rho_j}$, $d$ is the spatial dimension, $\Omega_{\mathrm{BZ}}$ is the area of the first Brillouin zone, and the static susceptibility is ${\chi := \frac{1}{N}\sum_{i,j} \avg{(\rho_i-\rho)(\rho_j - \rho)} = \rho(1 - \rho)}$. (The above asymptotic form is derived in the supplementary~\cite{supplemental}).
The classical-to-quantum mapping is simply to regard the rate matrix $H$ as a Hamiltonian expressed in terms of spin $S = 1/2$ operators~\cite{stinchcombe2001stochastic}, with real time evolution in the classical theory corresponding to imaginary time evolution in the quantum theory -- the (near-)equilibrium hydrodynamics of the classical model can thus be deduced from the low-energy properties of the spin Hamiltonian. Identifying $\ket{\uparrow}$ with the particles, and $\ket{\downarrow}$ with the holes, we have: 
\eqa{
&\bullet\;\mathrm{constraint\;that}\;i\;\mathrm{must\;be\;unoccupied} \iff S - \hS^z_i \nn \\
&\bullet\;\mathrm{constraint\;that}\;i\;\mathrm{must\;be\;occupied} \iff S + \hS^z_i
\nn \\
&\bullet\;\mathrm{hopping\;between}\;j\;\&\;l \iff S^2 - \hbS_j\cdot\hbS_l,
\label{eq:classical_to_quantum_mapping}
}
where $S = 1/2$ in the mapping from the classical model, but we will leave it as $S$ to organise the spin-wave expansion.
The spin Hamiltonian that corresponds to the constrained hopping on triangles shown in Fig.~\ref{fig:triangle_hopping_rules} is
\eqn{
\hH = \frac{1}{4 S^2}\sum_{ijl} \frac{g_{ijl}}{2} (S - \hS^z_i)(S^2 - \hbS_j\cdot\hbS_l),
\label{eq:triangular_chain}
}
where $g_{ijl} = 1$ if $i$, $j$, and $l$ are sites of a triangle, and $g_{ijl} = 0$ otherwise 
\footnote{The factor $1/2$ removes the double-counting from summing over all permutations.}.
We have divided the Hamiltonian by $4S^2$ ($=1$, a choice of timescale), such that the quadratic (non-interacting) term in the spin-wave expansion is $\mO(1)$. For any given particle density, the state
\eqn{
\ket{\Psi(\rho)} = \left(\sqrt{\rho}\,\ket{\uparrow} + \sqrt{1 - \rho}\,\ket{\downarrow}\right)^{\otimes N}
}
is an exact ground state of $\hH$~\footnote{Of course, this is a grand-canonical state, but it has density $\rho$ with probability $1$ in the thermodynamic limit; at finite size, it is a weighted sum of the ground states in different particle number sectors}. It may be checked explicitly that $\hH\ket{\Psi(\rho)} = 0$, and that
$\hH$ is itself positive semi-definite.
In particular, the ground state of the spin model indeed corresponds to the equilibrium state of the classical KCHP, where the particles ($\uparrow$-spins) are uniformly-randomly distributed over the sites.

\vspace{0.2cm}
\sectionn{Spin-wave expansion}
To analyse the low-energy modes, we consider the spin-wave expansion of Eq.~\eqref{eq:triangular_chain}, where we choose the quantisation axis as a function of $\rho$ such that $\ket{\Psi(\rho)}$ is the vacuum of the Holstein-Primakoff bosons. That is, the physical operators $S^{\mu}$
are given by
\eqa{
S^z = \cos\theta\,\tilde{S}^z - \sin\theta\,\tilde{S}^x, \;\;\;
S^x = \sin\theta\,\tilde{S}^z + \cos\theta\,\tilde{S}^x 
}
(and $S^y = \tilde{S}^y$), the mapping to the magnons is
\eqa{
\tilde{S}^z = S - \had\ha, \;\;\; 
\tilde{S}^{+} = \sqrt{2 S-\had\ha}\,\ha, \;\;\;
\tilde{S}^{-} = ({\tilde{S}^{+}})^{\dagger},
}
and $\cos\theta = 2\rho - 1$, $\sin\theta = 2\sqrt{\rho(1 - \rho)}$. Putting these into the Hamiltonian \eqref{eq:triangular_chain}, the first few terms in the $1/S$-expansion are
\eqa{
&\hH = \hH^{(2)} + \hH^{(3)} + \hH^{(4)} + ... \nn \\[0.1cm]
&\hH^{(2)} = \sum_k \epsilon(k) \had_k\ha_k \sim \sum_k Dk^2 \had_k\ha_k, \nn \\
&\hH^{(3)} = \frac{1}{\sqrt{N}}\sum_{k,q} f(k; q)\,\had_{\frac{k+q}{2}}\had_{\frac{k-q}{2}}\ha_k + \mathrm{h.c.} \nn \\
&\hH^{(4)} = \frac{1}{N}\sum_{k,q,q'} g(k; q, q')\,\had_{\frac{k+q'}{2}}\had_{\frac{k-q'}{2}}\ha_{\frac{k+q}{2}}\ha_{\frac{k-q}{2}}.
\label{eq:H_HP}
}
For the triangular chain, $D_{\mathrm{TC}} = 3(1 - \rho)$; for the triangular lattice, $D_{\mathrm{TL}} = \frac{3}{2}(1 - \rho)$.

As we are interested primarily in the long-wavelength properties of the system, we may similarly expand the vertices at small momentum. In the triangular chain, we have
\eqn{
f(k; q)
\;=\;
\begin{gathered}
\tik{
\draw[backarrow, thick] (0, 0) -- (0.5, 0);
\draw[forwardarrow, thick] (0, 0) -- (-0.25, 0.433);
\draw[forwardarrow, thick] (0, 0) -- (-0.25, -0.433);
\fill[black] (0, 0) circle [radius = 2pt];
\draw (-0.25, 0.433) node[left] {$\frac{k+q}{2}$};
\draw (-0.25, -0.433) node[left] {$\frac{k-q}{2}$};
\draw (0.5, 0) node[right] {$k$};
}
\end{gathered}
\;=\;
\frac{3\sqrt{\rho(1 - \rho)}}{2\sqrt{2 S}}\;\!k^2 + ...
\label{eq:cubic_vertex}
}
and
\eqa{
g(k;\, &q, q')
\;=\;
\begin{gathered}
\tik{
\draw[forwardarrow, thick] (0, 0) -- (-0.5, -0.5);
\draw[forwardarrow, thick] (0, 0) -- (-0.5, 0.5);
\draw[backarrow, thick] (0, 0) -- (0.5, -0.5);
\draw[backarrow, thick] (0, 0) -- (0.5, 0.5);
\draw (-0.5,0.4) node[left]{$\frac{k+q'}{2}$};
\draw (-0.5,-0.4) node[left]{$\frac{k-q'}{2}$};
\draw (0.5,0.4) node[right]{$\frac{k+q}{2}$};
\draw (0.5,-0.4) node[right]{$\frac{k-q}{2}$};
\fill[black] (0,0) circle [radius=2pt];
}
\end{gathered}\nn \\
\;&=\;
-\;\!\frac{3(2 - 3\rho)}{8 S}\;\!k^2 + \frac{3(1 - \rho)}{16 S}\;\!(q^2 + q'\;\!^2) + ...\,.
\label{eq:quartic_vertex}
}
A more detailed derivation of the spin-wave expansion, and the general full forms of $f$ and $g$, are given in the supplementary~\cite{supplemental}.

\vspace{0.2cm}
\sectionn{Continuum limit \& effective field theory}
In the same spirit as expanding at small momentum, we may take the continuum limit and cast the model in the path integral formalism as an effective field theory (EFT), with the (imaginary-time) Lagrangian
\eqa{
\mL \,=\, \barphi&\;\!\pd_t\phi - D\;\!\barphi\;\!\nabla^2\phi - \lambda_{(3)}\left(\barphi\;\!\barphi\;\!\nabla^2\phi + \mathrm{h.c.}\right) \nn \\
&- (\lambda_{(4)} + 2\gamma_{(4)})\left(\barphi\;\!\barphi\;\!\phi\;\!\nabla^2\phi + \mathrm{h.c.}\right) \nn \\
&- (\lambda_{(4)} - 2\gamma_{(4)})\left(\barphi\;\!\barphi\;\!\nabla\phi\cdot\nabla\phi + \mathrm{h.c.}\right) + ...\,,
}
where $\phi(t, x)$ is a (complex) bosonic field. For the triangular chain, the bare value of $\lambda_{(3)}$ is the coefficient of $k^2$ in Eq.~\eqref{eq:cubic_vertex}, and the bare values of $\lambda_{(4)}$ and $\gamma_{(4)}$ are the coefficients of $k^2$ and $q^2 + q'\,\!^2$, respectively, in Eq.~\eqref{eq:quartic_vertex}. The couplings $\lambda_{(n)},\gamma_{(n)}$ etc.\ are $\mO(S^{-(n-2)/2})$. 

The appearance of a small parameter $1/S$ is where our method differs from other field-theoretic approaches to stochastic processes, such as Doi-Peliti~\cite{doi1976stochastic,peliti1985path,grassberger1980fock}.
With such methods, one needs often to resort immediately to an RG analysis~\cite{cardy1996theory,cardy1998field} using dimensional regularisation or other advanced techniques~\cite{tauber2005applications,canet2004nonperturbative}.

The EFT has a natural momentum cutoff ${\Lambda \sim 1/a}$, where $a$ is the lattice spacing, in powers of which we express the scaling dimensions. Indeed, for a diffusive free theory, ${[\mL] = 3}$ (${[d\tau] = -2}$, ${[dx] = -1}$), whence follows ${[\barphi]=[\phi] = 1/2}$, and thence ${[D] = 0}$, ${[\lambda_{(n)}] = -(n-2)/2}$~\footnote{In $d$ spatial dimensions, $[\mL] = 2 + d\Rightarrow[\phi]=d/2$, $[D] = 0$, and ${[\lambda_{(n)}] = -d(n-2)/2}$.}. There are also couplings for higher derivative terms, $\barphi(\nabla^2)^2\phi, \barphi\barphi\,\nabla^2\phi\,\nabla^2\phi$,\hspace{0.1cm} $\barphi\,\nabla^2\barphi\,\nabla^2\phi\,\nabla^2\phi$, etc.; but they are more irrelevant, as their engineering dimensions are even more negative. 
%
%
%
%
\begin{figure}
    \centering
    \tik{
    \draw[dotted, thin] (-0.2,1.8) -- (1,1.8) -- (1.8,1.8) -- (3.0,1.8);
    \draw[dotted, thin] (-0.2,1.2) -- (1,1.2) -- (1.8,1.2) -- (3.0,1.2);
    \draw[dotted, thin] (-0.2,0.6) -- (1,0.6) -- (1.8,0.6) -- (3.0,0.6);
    \draw[dotted, thin] (-0.2,0.0) -- (1,0.0) -- (1.8,0.0) -- (3.0,0.0);
    \fill[white] (0.2,1.8) circle [radius=4pt];
    \draw[black] (0.2,1.8) circle [radius=4pt];
    \fill[gray] (1,1.8) circle [radius=4pt];
    \draw[black] (1,1.8) circle [radius=4pt];
    \fill[white] (1.8,1.8) circle [radius=4pt];
    \draw[black] (1.8,1.8) circle [radius=4pt];
    \draw[<->, thick] (1.14, 1.9) arc (150:30:0.3);
    \fill[white] (2.6,1.8) circle [radius=4pt];
    \draw[black] (2.6,1.8) circle [radius=4pt];
    \fill[white] (0.2,1.2) circle [radius=4pt];
    \draw[black] (0.2,1.2) circle [radius=4pt];
    \fill[gray] (1,1.2) circle [radius=4pt];
    \draw[black] (1,1.2) circle [radius=4pt];
    \fill[white] (1.8,1.2) circle [radius=4pt];
    \draw[black] (1.8,1.2) circle [radius=4pt];
    \draw[<->, thick] (1.14, 1.3) arc (150:30:0.3);
    \fill[gray] (2.6,1.2) circle [radius=4pt];
    \draw[black] (2.6,1.2) circle [radius=4pt];
    \fill[gray] (0.2,0.6) circle [radius=4pt];
    \draw[black] (0.2,0.6) circle [radius=4pt];
    \fill[gray] (1,0.6) circle [radius=4pt];
    \draw[black] (1,0.6) circle [radius=4pt];
    \fill[white] (1.8,0.6) circle [radius=4pt];
    \draw[black] (1.8,0.6) circle [radius=4pt];
    \draw[<->, thick] (1.14, 0.7) arc (150:30:0.3);
    \fill[white] (2.6,0.6) circle [radius=4pt];
    \draw[black] (2.6,0.6) circle [radius=4pt];
    \fill[gray] (0.2,0.0) circle [radius=4pt];
    \draw[black] (0.2,0.0) circle [radius=4pt];
    \fill[gray] (1,0.0) circle [radius=4pt];
    \draw[black] (1,0.0) circle [radius=4pt];
    \fill[white] (1.8,0.0) circle [radius=4pt];
    \draw[black] (1.8,0.0) circle [radius=4pt];
    \draw[<->, thick] (1.14, 0.1) arc (150:30:0.3);
    \fill[gray] (2.6,0.0) circle [radius=4pt];
    \draw[black] (2.6,0.0) circle [radius=4pt];
    \draw (1.5, 2.2) node[above]{Four-site gate hopping rules};
    \draw (3.5, 1.8) node[right]{$\Gamma_{0,0}\;\;= $};
    \draw (3.5, 1.2) node[right]{$\Gamma_{0,1}\;\;= $};
    \draw (3.5, 0.6) node[right]{$\Gamma_{1,0}\;\;= $};
    \draw (3.5, 0.0) node[right]{$\Gamma_{1,1}\;\;= $};
    \draw (5.2, 2.2) node[above]{OR};
    \draw (5.2, 1.6) node[above]{$0$};
    \draw (5.2, 1.0) node[above]{$1$};
    \draw (5.2, 0.4) node[above]{$1$};
    \draw (5.2, -0.2) node[above]{$1$};
    \draw (6.2, 2.2) node[above]{XOR};
    \draw (6.2, 1.6) node[above]{$0$};
    \draw (6.2, 1.0) node[above]{$1$};
    \draw (6.2, 0.4) node[above]{$1$};
    \draw (6.2, -0.2) node[above]{$0$};
    \draw (7.2, 2.2) node[above]{XNOR};
    \draw (7.2, 1.6) node[above]{$1$};
    \draw (7.2, 1.0) node[above]{$0$};
    \draw (7.2, 0.4) node[above]{$0$};
    \draw (7.2, -0.2) node[above]{$1$};
    }
    \caption{Summary of the hopping rules for the four-site gate chains, and the values of the model parameters for the three chains we consider. The OR chain is also known as the GLT chain~\cite{singh2021subdiffusion}. ${\Gamma_{ab} = 1}$ (${\Gamma_{ab} = 0}$) means that hopping on a bond is allowed (disallowed) when the occupancies of the sites to the left and right of the bond are $a$ and $b$, respectively.
    } 
    \label{fig:4-site_gate_hopping_rules}
\end{figure}
%
\begin{figure*}
    \centering
    \includegraphics[width=\linewidth]{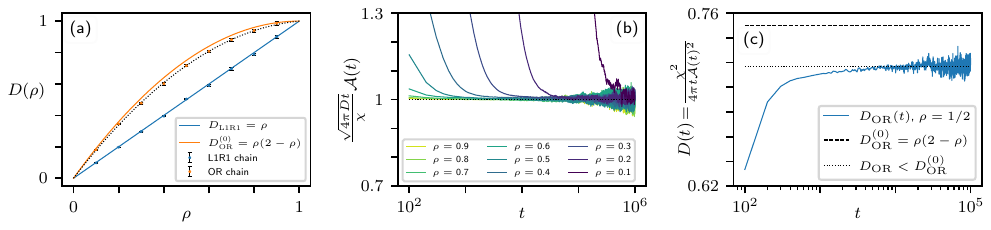}
    \caption{
    Corrections to the bare diffusion constant in the OR chain, a four-site gate model. (a) Numerical estimates of the diffusion constant $D(\rho)$ for the OR chain (four-site gates) and the L1R1 chain (three-site gates). Any hopping process allowed in one model is allowed in the other (though with different rates), but only the OR chain has its diffusion constant corrected. (b) Convergence of the autocorrelator in the L1R1 chain to the non-interacting spin wave prediction $D_{\mathrm{L1R1}} = \rho$. (c) Time-dependent estimate of the diffusion constant at $\rho = 1/2$ in the OR chain, which clearly converges to a value less than $D^{(0)}_{\mathrm{OR}}(\rho = 1/2) = 3/4$. $N = 8192$ with PBCs and $10^4$ states in the ensemble in all cases.
    }
    \label{fig:4-site_autocorrs}
\end{figure*}
%
This is an effective field theory in the sense of Weinberg~\cite{georgi1993effective,burgess2007introduction}, and, at long-wavelength ($k \ll \Lambda$), is a valid replacement for the original lattice model if the coefficients are chosen properly~\cite{michailidis2024corrections}. 

Now, in principle, the RG flow of this Lagrangian could generate apparently relevant terms with no derivatives such as $\barphi\;\!\barphi\;\!\phi$ and $\barphi\;\!\barphi\;\!\phi\;\!\phi$. However, these terms are forbidden by the degeneracy of the ground state manifold $\ket{\Psi(\rho)}$; and all terms in the effective Lagrangian must appear with at least two powers of $\nabla$ (or one of $\pd_t$).

\vspace{0.2cm}
\sectionn{Non-renormalisation of the diffusion constant}
We now argue that non-interacting spin-wave theory gives the \textit{exact} diffusion constant for the triangular model in any dimension. By construction, $\ket{\Psi(\rho)}$ is the vacuum of the magnons $\ha$; crucially, it is also an exact ground state.
In the perturbative expansion, this causes a large subset of diagrams to vanish, as any closed (respecting the arrows) loop of propagators is proportional to a normal-ordered vacuum expectation value. 

Any contribution to the magnon self-energy must, therefore, start with the $1\to2$ vertex and end with the $2\to1$ vertex. The full one-magnon Green's function, then, can be expressed implicitly as
\eqa{
\begin{gathered}
\tik{
\draw[backarrow, double] (0, 0) -- (1.2, 0);
}
\end{gathered}
\;=\;
\begin{gathered}
\tik{
\draw[backarrow, thick] (0, 0) -- (1.2, 0);
}
\end{gathered}
\;+\;
\begin{gathered}
\tik{
\draw[forwardarrow, thick] (0.0, 0) -- (-0.7, 0);
\draw[forwardarrow, double] (1.7, 0) -- (1.18, 0);
\draw[backarrow, thick] (0,0) arc (180:0:0.5);
\draw[backarrow, thick] (0,0) arc (-180:0:0.5);
\draw[fill=gray] (0.95, 0.24) -- (1.2, 0.0) -- (0.95, -0.24) -- cycle;
\fill[black] (0,0) circle [radius=2pt];
}
\end{gathered},
\label{eq:Green's_function}
}
where all two-magnon scattering processes have been absorbed as corrections to the $1\to2$ vertex. Within the EFT, however, all these corrections simply result in a renormalisation of the coefficients in the expansion of $f(k; q) \to \tilde{f}(k; q)$. We know that $\tilde{f}(k; q)$ must remain even in both $k$ (by parity) and $q$ (by exchange); and, as noted above, no constants or terms $\sim q^2$ can appear in the expansion because they would generate a mass term. Thus, up to second order in momentum, the effect of all these corrections is simply to renormalise $\lambda_{(3)} \to \tilde{\lambda}_{(3)}$.
Evaluating the diagram, then, we find the self-energy has the form
\eqn{
\Sigma(\omega, k) = \lambda_{(3)}\tilde{\lambda}_{(3)} k^4 \mF(\omega,k) + ...\,,
\label{eq:self-energy}
}
where ${\mF(\omega, k) \sim 1/\sqrt{Dk^2 - 2i\omega}}$ in one dimension, and ${\mF(\omega, k) \sim \log(Dk^2 - 2i\omega)}$ in two dimensions. The pole of the Green's function is thus not shifted at $\mO(k^2)$, so the observed diffusion constant does not change (the explicit evaluation of the diagrams, and some subtleties of complex analysis, are given in the supplementary~\cite{supplemental}).

Although we have motivated our analysis with the triangular models \eqref{eq:triangular_chain}, the conclusion that the bare diffusion constant is exact depends only on the facts that (i) the cubic $1\leftrightarrow2$ vertices are $\mO(k^2)$, and (ii) that they are the only terms that directly connect the one-magnon sector to the multiple-magnon sectors. There is a general class of KCHPs with these properties, which are realised if the model (i) is parity-symmetric~\footnote{If parity symmetry is absent, the cubic vertex $f(k; q)$ generically contains terms $\sim ikq^2$, which will generate terms of $\mO(k^2)$ in the self-energy. The (localised) $U(1)$-East model of Ref.~\cite{singh2021subdiffusion} is a sum of three-site gates, but is not parity symmetric.}, and (ii) can be written as a sum of three-site gates -- that is, hopping between two sites is controlled by the occupancy of a third. We conjecture that, for all such models, the exact diffusion constant can be read off from the bare magnon dispersion, regardless of dimension or lattice geometry.

It is also useful and important to compare with quantum ferromagnets (FMs). The simplest examples have exact product ground states~\cite{feynman2018statistical,ceulemans2001finite}, with vanishing zero-point fluctuations and exact single-magnon eigenstates, despite being energy-matched to the multiple-magnon continua~\cite{ceulemans2001finite}.  This is highly non-generic~\cite{beekman2015criteria}, since Goldstone modes should generally have finite lifetimes except in the long-wavelength limit~\cite{chaikin1995principles}; however, `spontaneous magnon decay' at $T=0$ can and does exist in more general FM models (and materials)~\cite{zhitomirsky2013colloquium,stephanovich2011spontaneous}. Importantly, details of interactions matter -- e.g., whilst one might expect analytic self-energies $\Sigma \sim k^4$~\cite{chaikin1995principles}, the result is $\Sigma \sim |k|^5$ in planar FMs~\cite{stephanovich2011spontaneous}, and the KCHP on the triangular lattice has $\Sigma \sim k^4 \log |k|$ in 2D, and $\sim |k|^3$ in 1D.

\vspace{0.2cm}
\sectionn{Numerical results}
We perform numerical simulations to test these conclusions, on the triangular chain and the triangular lattice (cf. Fig.~\ref{fig:triangle_hopping_rules}). For each density $\rho$, we construct an ensemble of $10^4$ (grand-canonical) initial states, placing a particle on each site with probability $\rho$. For each state in the ensemble, we construct a trajectory, where one timestep consists of choosing a random triangle $N$ times and applying the hopping rules. The numerical estimate of the diffusion constant is extracted from the autocorrelator~\eqref{eq:autocorrelator}. Results for the triangular chain and triangular lattice models are shown in Fig.~\ref{fig:triangular_diffusion}; the observed diffusion constants are consistent with the non-interacting spin wave predictions at all densities $\rho$.

\vspace{0.2cm}

\sectionn{Four-site gate models}
We now address how the above arguments for the non-renormalisation of the diffusion coefficient can break down if the dynamics is given by \textit{four}-site gates.
%
Specifically, consider a 1D chain, where hopping between neighbouring sites is conditioned on the occupancies of the sites to the left and right of that bond. These hopping rules are illustrated in Fig.~\ref{fig:4-site_gate_hopping_rules}. Using the classical-to-quantum mapping, we can write a corresponding spin Hamiltonian as
\eqa{
\hH = \sum_{a,b\in \{0, 1\}}  \frac{\Gamma_{ab}}{8 S^3}\sum_j \Bigl[(S - \sigma_a&\hS^z_{j-1})(S^2 - \hbS_{j}\cdot\hbS_{j+1}) \nn \\[-0.35cm]
&\times(S - \sigma_b\hS^z_{j+2})\Bigr],
\label{eq:4_site_H}
}
where ${\sigma_0 = +1}$, ${\sigma_1 = -1}$, and ${\Gamma_{ab} = 1}$ (${\Gamma_{ab} = 0}$) means that hopping on a bond is allowed (disallowed) when the occupancies of the sites to the left and right of the bond are $a$ and $b$, respectively. For these models, non-interacting spin-wave theory predicts
\eqn{
D^{(0)} = \frac{\Gamma_{0,0}}{2}(1-\rho)^2 + \frac{\Gamma_{0,1} + \Gamma_{1,0}}{2}\rho(1 - \rho) + \frac{\Gamma_{1,1}}{2}\rho^2.
}
Notably, $\ket{\Psi(\rho)}$ is still an exact ground state~\footnote{In fact, it is an exact ground state for the spin Hamiltonian corresponding to \textit{any} number-conserving, translationally-invariant hopping process.}, and so any diagram with a closed propagator loop still vanishes. What is different for four-site gates is that, in the spin-wave expansion of Eq.~\eqref{eq:4_site_H}, there is an additional term at $\mO(1/S)$ which does not appear in Eq.~\eqref{eq:H_HP}, to wit
\eqn{
\hH^{(4)'} = \frac{1}{N}\sum_{k,p,q} h(k; q, q')\;\!\had_{\frac{k+q-q'}{3}}\;\!\had_{\frac{k-2q}{3}}\;\!\had_{\frac{k+q+q'}{3}}\;\!\ha_k + \mathrm{h.c.},
}
corresponding to the direct decay of one magnon into three. Diagrammatically, the existence of the $1\leftrightarrow3$ vertices generates new terms in the self-energy, in particular
\eqn{
\Sigma_{1\to3\to1}(\omega, k)
\;=\;
\begin{gathered}
\tik{
\draw[backarrow, thick] (0,0) arc (180:0:0.5);
\draw[backarrow, thick] (0,0) arc (-180:0:0.5);
\draw[backarrow, thick] (0, 0) -- (1.0, 0);
\draw[thick] (0, 0) -- (-0.2, 0);
\draw[thick] (1.2, 0) -- (1.0, 0);
\fill[black] (0,0) circle [radius=2pt];
\fill[black] (1.0,0) circle [radius=2pt];
}
\end{gathered}
\;=\; \delta D \;\!k^2 + \mO(k^3),
}
which shifts the diffusion pole (at this order of perturbation theory) to $D^{(0)} - \delta D$. The crucial term in $h(k, q, q')$ occurs at fourth order in momentum, ${\sim \zeta_{(4)}kq(q-q')(q+q')}$, but has only one power of the external momentum $k$~\footnote{This cannot happen in the higher derivative terms of $f(k; q)$ and $g(k; q, q')$, since they are even in the external momentum $k$.}. The bare value of $\zeta_{(4)}$ is
\eqn{
\zeta_{(4)} = -\frac{20}{81}\frac{\rho(1-\rho)}{4S}\sum_{a,b}\Gamma_{ab}\sigma_a\sigma_b,
}
and $\delta D = \sqrt{3}\,\zeta_{(4)}^2\Lambda^6/(1024\pi D^{(0)}) > 0$. (We give the full form of $h$ and explicitly evaluate the diagram in the supplementary~\cite{supplemental}). This is a small correction in the OR chain, and, numerically, we find that the diffusion constant of the OR chain indeed suffers a weak negative correction (Fig.~\ref{fig:4-site_autocorrs}). It is a stronger correction for the XNOR and XOR chains -- and in those models, in fact, the diffusion constant is renormalised all the way to zero and they evince subdiffusion instead~\cite{singh2021subdiffusion} (we present numerical evidence for that in the supplementary~\cite{supplemental}), though our present analysis does not allow us to predict their subdiffusive exponent $z = 4$.

Finally, as a further test of our conjecture that any parity-symmetric, three-site gate KCHP has its exact diffusion constant given by non-interacting spin wave theory, consider the `L1R1' chain~\footnote{which we name thus because it is the sum of an `L1' gate and an `R1' gate, where hopping is allowed if the site on the left/right of the three is occupied (`1').}, with the following hopping rules: pick a random bond, then randomly select the site to the left or right of the bond (with equal probability); a hop occurs if that site is occupied. 

The L1R1 chain is almost identical to the OR chain -- indeed, a hop is possible in one if and only if it is possible in the other. The rates, however, are slightly different, and the L1R1 chain is a sum of \textit{three}-site gates,
\eqa{
\hH_{\mathrm{L1R1}} = \frac{1}{4 S^2}\sum_j &(S + \hS^z_{j-1})(S^2 - \hbS_j\cdot\hbS_{j+1}) \nn \\[-0.2cm]
&+ (S^2 - \hbS_j\cdot\hbS_{j+1})(S + \hS^z_{j+2}).
}
Non-interacting spin-wave theory predicts $D_{\mathrm{L1R1}} = \rho$, and we show in Fig.~\ref{fig:4-site_autocorrs} that the numerics are consistent with this being exact.

\vspace{0.2cm}
\sectionn{Discussion \& conclusions}
%
In this Letter, we have used a classical-to-quantum correspondence for the rate matrix of KCHPs to show how their hydrodynamics can be obtained from a semi-classical $1/S$ expansion of the quantum theory; the dominant terms are (approximately) independent hydrodynamic modes, and the higher order terms introduce interactions between them. 
In the long-wavelength limit we obtain a continuum EFT, with which we show that for the class of KCHPs with parity-symmetry and three-site gate dynamics -- and regardless of lattice geometry or spatial dimension -- the interactions do not renormalise the diffusion constant, which is thus obtained exactly from non-interacting spin-wave theory; and we have supported this claim with extensive numerical simulation of the KCHPs. The fact that there are corrections to the diffusion constant of the four-site gate OR chain, but no corrections in the three-site gate L1R1 chain -- despite all hops allowed in one being allowed in the other -- is particularly striking.

Our method points the way towards an analytic attack on KCHPs with the full artillery of effective field theory; an obvious direction for future work in this framework is to capture the subdiffusive regimes of KCHPs such as the XNOR and XOR chains, with the interactions driving the system away from the diffusive $k^2$-Gaussian fixed point towards a subdiffusive $k^{z > 2}$-Gaussian fixed point; or, in some models, even a non-Gaussian fixed point.

\sectionn{Acknowledgements}
AS would like to thank Giovanni Villadoro and Sergio Benvenuti for enlightening discussions on the uses of EFT. AMcR acknowledges discussions during the Student Workshop on Integrability at ELTE, Budapest.  VO thanks Daniel P. Arovas, Luca Delacretaz, Paul Krapivsky, David A. Huse, Paolo Glorioso, Sarang Gopalakrishnan, Abhishek Raj for useful physics discussions and the latter three for joint work on a related project. The work of AS and AMcR was funded by the European Union--NextGenerationEU under the project NRRP ``National Centre for HPC, Big Data and Quantum Computing (HPC)'' CN00000013 (CUP D43C22001240001) [MUR Decree n.\ 341--15/03/2022] -- Cascade Call launched by SPOKE 10 POLIMI: ``CQEB'' project, and from the National Recovery and Resilience Plan (NRRP), Mission 4 Component 2 Investment 1.3 funded by the European Union NextGenerationEU, National Quantum Science and Technology Institute (NQSTI), PE00000023, Concession Decree No. 1564 of 11.10.2022 adopted by the Italian Ministry of Research, CUP J97G22000390007.




\bibliography{refs} 

\onecolumngrid

  \cleardoublepage
  \begin{center}
    \textbf{\large Supplementary material}
  \end{center}
\setcounter{equation}{0}
\setcounter{figure}{0}
\setcounter{table}{0}
\makeatletter
\renewcommand{\theequation}{S\arabic{equation}}
\renewcommand{\thefigure}{S\arabic{figure}}
\renewcommand{\thetable}{S\arabic{table}}
\setcounter{section}{0}
\renewcommand{\thesection}{S-\Roman{section}}

\setcounter{secnumdepth}{2}

\section*{Contents}

In this supplementary material we give the details of the classical-to-quantum mapping and the spin wave expansion for the triangular chain and triangular lattice models; derive the asymptotic form of the autocorrelation function; discuss some complex-analytic subtleties of the full (interacting) magnon Green's function and check the self-energy up to $\mO(1/S^2)$; explicitly evaluate one of the diagrams that can reduce the diffusion constant in the four-site gate models; and present numerical evidence that the XOR and XNOR chains are subdiffusive at almost all densities.

\section{Classical-to-quantum mapping and spin-wave expansion}

The spin-wave expansion we employ in the main text is simply the standard Holstein-Primakoff expansion; but for the convenience of the reader, we spell out most of the details here. 

Cf. Fig.~1, a single step of the kinetically-constrained hopping process for the triangular chain and triangular lattice models consists of choosing a random \textit{triangle} (not a random particle), and, if it contains exactly one particle, that particle hops to one of the other two sites with equal probability. The rate matrix for this process on those three sites, say, $i$, $j$, $l$, is, in the canonical basis $\ket{\uparrow\uparrow\uparrow}$, $\ket{\uparrow\uparrow\downarrow}$, ..., $\ket{\downarrow\downarrow\uparrow}$, $\ket{\downarrow\downarrow\downarrow}$,
\eqn{
H_{ijl} = \begin{pmatrix}
    0 & 0 & 0 & 0 & 0 & 0 & 0 & 0 \\
    0 & 0 & 0 & 0 & 0 & 0 & 0 & 0 \\
    0 & 0 & 0 & 0 & 0 & 0 & 0 & 0 \\
    0 & 0 & 0 & -1 & 0 & \frac{1}{2} & \frac{1}{2} & 0 \\
    0 & 0 & 0 & 0 & 0 & 0 & 0 & 0 \\
    0 & 0 & 0 & \frac{1}{2} & 0 & -1 & \frac{1}{2} & 0 \\
    0 & 0 & 0 & \frac{1}{2} & 0 & \frac{1}{2} & -1 & 0 \\
    0 & 0 & 0 & 0 & 0 & 0 & 0 & 0
\end{pmatrix}
=
(S - S_i^z)(S^2 - \hbS_j\cdot\hbS_l) + (S - S_j^z)(S^2 - \hbS_l\cdot\hbS_i) + (S - S_l^z)(S^2 - \hbS_i\cdot\hbS_j),
}
where, as mentioned in the main text, we identify $\ket{\uparrow} \cong \ket{1}$ with the classical particles, and $\ket{\downarrow} \cong \ket{0}$ with the holes.

A single \textit{timestep} of the kinetically-constrained hopping process for a system of $N$ sites consists of $N$ independent steps, where a step consists of choosing a random triangle and applying the hopping rules (the same triangle may be selected more than once in the sequence).

Identifying the sum of all the rate matrices as a Hamiltonian, we may write this generally as
\eqn{
\hH = \frac{1}{4S^2}\sum_{i,j,l} \frac{g_{ijl}}{2} (S - \hS^z_i)(S^2 - \hbS_j\cdot\hbS_l),
}
where $g_{ijl} = 1$ if $i$, $j$, and $l$ are sites of a triangle, and $g_{ijl} = 0$ otherwise (and the factor $1/2$ removes the double-counting from summing over all permutations). As discussed in the main text, the crucial observation is that $\hH$ is positive-semidefinite, and that
\eqn{
\hH\ket{\Psi(\rho)} = \hH\left(\sqrt{\rho}\,\ket{\uparrow} + \sqrt{1 - \rho}\,\ket{\downarrow}\right)^{\otimes N} = 0.
}
Thus, for all $\rho$, $\ket{\Psi(\rho)}$ is an exact ground state, and has density $\rho$ with probability one in the limit $N\to\infty$.

To investigate the hydrodynamics of the classical hopping process at density $\rho$ -- which corresponds to the low-energy properties of $\hH$ -- we construct the spin wave expansion above the corresponding $\ket{\Psi(\rho)}$. That is, we define rotated spin operators $\tilde{S}^x$, $\tilde{S}^z$ such that
\eqn{
S^x = \sin\theta\,\tilde{S}^z + \cos\theta\,\tilde{S}^x,\;\;\;
S^y = \tilde{S}^y,
\;\;\;
S^z = \cos\theta\,\tilde{S}^z - \sin\theta\,\tilde{S}^x,
}
and $\ket{\uparrow_{\tilde{z}}}^{\otimes N} = \ket{\Psi(\rho)}$. That is, $\cos\theta = 2\rho - 1$, $\sin\theta = 2\sqrt{\rho(1 - \rho)}$, and the bosons are then defined via the standard mapping,
\eqa{
\tilde{S}^z = S - \had\ha, \;\;\; 
\tilde{S}^{+} = \sqrt{2S-\had\ha}\,\ha, \;\;\;
\tilde{S}^{-} = ({\tilde{S}^{+}})^{\dagger}.
}
In terms of the bosons, then, the factors in the terms of the spin Hamiltonian are
\eqn{
S - \hS^z_i = S(1 - \cos\theta) +\cos\theta\had_i\ha_i + \frac{\sin\theta}{2}\had_i\sqrt{2S - \had_i\ha_i} + \frac{\sin\theta}{2}\sqrt{2S - \had_i\ha_i}\ha_i,
\label{eq:spin_wave_constraint_factor}
}
and
\eqn{
S^2 - \hbS_j\cdot\hbS_l = S\had_j\ha_j + S\had_l\ha_l - \had_j\ha_j\had_l\ha_l - \frac{1}{2}\had_j\sqrt{2S - \had_j\ha_j}\sqrt{2S - \had_l\ha_l}\ha_l - \frac{1}{2}\had_l\sqrt{2S - \had_l\ha_l}\sqrt{2S - \had_j\ha_j}\ha_j.
\label{eq:spin_wave_hopping_factor}
}
We now expand the Hamiltonian $\hH = \hH^{(2)} + \hH^{(3)} + \hH^{(4)} + \,...$ in powers of $1/S$ in the usual way by expanding the square roots. We have the quadratic terms
\eqn{
\hH^{(2)} = \frac{1 - \cos\theta}{4}\sum_{ijl} \frac{g_{ijl}}{2}\left(\had_j\ha_j + \had_l\ha_l - \had_j\ha_l - \had_l\ha_j\right),
}
from which the bare diffusion constant is obtained. The cubic terms are
\eqn{
\hH^{(3)} = \frac{\sin\theta}{4\sqrt{2S}}\sum_{ijl} \frac{g_{ijl}}{2}\left(\had_i + \ha_i\right)\left(\had_j\ha_j + \had_l\ha_l - \had_j\ha_l - \had_l\ha_j\right),
}
and the quartic terms are
\eqa{
\hH^{(4)} = \;&\frac{1 - \cos\theta}{4S}\sum_{ijl}\frac{g_{ijl}}{2}\left(- \had_j\ha_j\had_l\ha_l 
+ \frac{1}{4}\had_j\had_j\ha_j\ha_l 
+ \frac{1}{4}\had_j\had_l\ha_l\ha_l
+ \frac{1}{4}\had_l\had_j\ha_j\ha_j
+ \frac{1}{4}\had_l\had_l\ha_l\ha_j\right) \nn \\
&+ \frac{\cos\theta}{4S} \sum_{ijl}\frac{g_{ijl}}{2}\had_i\left(\had_j\ha_j + \had_l\ha_l - \had_j\ha_l - \had_l\ha_j\right)\ha_i.
}
There are quintic, etc., terms at higher order in $1/S$, though we will not give them explicitly. It is crucial to note, however, that \textit{at no order} in $1/S$ are there any terms that change the boson number by more than one. This is easy to see by inspecting Eqs.~\eqref{eq:spin_wave_constraint_factor} \& \eqref{eq:spin_wave_hopping_factor}, since the square root factors conserve the total boson number; it is this fact that enables us to write down a diagrammatically-simple Dyson equation \eqref{eq:Green's_function_sup} for the full Green's function.

Returning to the terms that we do consider, it remains only to take advantage of the translational symmetry and transform to momentum space. For the Fourier transform conventions, we define
\eqn{
\ha_j = \frac{1}{\sqrt{N}} \sum_{\bk} e^{i \bk\cdot \br_j} \ha_{\bk}.
}

The non-interacting (quadratic) terms in momentum space are, therefore,
\eqn{
\hH^{(2)} = \sum_{\bk} \underbrace{\frac{1 - \cos\theta}{2}}_{1 - \rho} \left(\frac{1}{N}\sum_{ijl} \frac{g_{ijl}}{2}\left(1 - \cos\left(\bk\cdot(\br_j - \br_l)\right)\right)\right) \had_{\bk}\ha_{\bk} =: \sum_{\bk} \epsilon(\bk)\had_{\bk}\ha_{\bk}.
}

For the triangular chain, each site $i$ belongs to three triangles, and the possible values of $j - l$ are $\pm2$ (which occurs twice) and $\pm1$ (which occurs four times, counting all permutations). This gives
\eqn{
\epsilon(k) = (1 - \rho)(3 - 2\cos k - \cos2k) \sim 3(1 - \rho)k^2 \;\;\;\Rightarrow\;\;\; D_{\mathrm{TC}} = 3(1 - \rho).
}

And for the triangular lattice, each site $i$ is part of six different triangles. The possible values of $\br_j - \br_l$ are $\pm\bx$, $\pm(\bx/2 + \sqrt{3}\by/2)$, and $\pm(\bx/2 - \sqrt{3}\by/2)$, each of which occurs four times including all perturbations, and we find
\eqa{
\epsilon(\bk) = (1 - \rho)&\left(6 - 2\cos k_x - 2 \cos\left(\frac{k_x}{2} + \frac{\sqrt{3}k_y}{2}\right) - 2 \cos\left(\frac{k_x}{2} - \frac{\sqrt{3}k_y}{2}\right)\right) \sim \frac{3}{2}(1 - \rho)\bk^2 \nn \\
&\;\;\Rightarrow\;\; D_{\mathrm{TL}} = \frac{3}{2}(1 - \rho).
}

Similarly, the cubic terms become 
\eqn{
\hH^{(3)} = \frac{1}{\sqrt{N}}\sum_{\bk,\bq} f(\bk; \bq)\,\had_{\frac{\bk+\bq}{2}}\had_{\frac{\bk-\bq}{2}}\ha_{\bk} + \mathrm{h.c.},
}
where we have used that translational invariance will impose momentum conservation, and
\eqa{
f(\bk; \bq) = \frac{\sin\theta}{4\sqrt{2S}}\frac{1}{N}\sum_{ijl}\frac{g_{ijl}}{2}\cdot\frac{1}{2}\Biggl[ e^{-i\frac{\bk+\bq}{2}\cdot\br_i}\left(e^{-i\frac{\bk-\bq}{2}\cdot\br_j + i\bk\cdot\br_j} 
+ e^{-i\frac{\bk-\bq}{2}\cdot\br_l + i\bk\cdot\br_l} 
- e^{-i\frac{\bk-\bq}{2}\cdot\br_j + i\bk\cdot\br_l} 
- e^{-i\frac{\bk-\bq}{2}\cdot\br_l + i\bk\cdot\br_j}\right) \nn \\
+ e^{-i\frac{\bk-\bq}{2}\cdot\br_i}\left(e^{-i\frac{\bk+\bq}{2}\cdot\br_j + i\bk\cdot\br_j} 
+ e^{-i\frac{\bk+\bq}{2}\cdot\br_l + i\bk\cdot\br_l} 
- e^{-i\frac{\bk+\bq}{2}\cdot\br_j + i\bk\cdot\br_l} 
- e^{-i\frac{\bk+\bq}{2}\cdot\br_l + i\bk\cdot\br_j}\right)\Biggr],
}
and we have used the fact that the creation operators commute amongst themselves to symmetrise between $\bq$ and $-\bq$. Expanding at small momentum, this becomes
\eqn{
f(\bk; \bq) \sim \frac{\sin\theta}{4\sqrt{2S}} \frac{1}{N}\sum_{ijl}\frac{g_{ijl}}{2}\,\frac{1}{2}\left(\bk\cdot(\br_j - \br_l)\right)^2.
}
Using the same considerations as for the quadratic terms as to the possible values of $\br_j - \br_l$, we find
\eqn{
f(k; q) \sim \frac{\sin\theta}{4\sqrt{2S}} \frac{1}{2}\cdot\frac{1}{2}\left(2(2k)^2 + 4(k)^2\right) = \frac{3\sin\theta}{4\sqrt{2S}}k^2 = \frac{3\sqrt{\rho(1-\rho)}}{2\sqrt{2S}}k^2
}
for the triangular chain, and
\eqn{
f(\bk; \bq) \sim \frac{\sin\theta}{4\sqrt{2S}} \frac{1}{2}\cdot\frac{1}{2}\left(4\left(k_x\right)^2 + 4\left(\frac{k_x}{2} + \frac{\sqrt{3}k_y}{2}\right)^2 + 4\left(\frac{k_x}{2} - \frac{\sqrt{3}k_y}{2}\right)^2\right) = \frac{3\sin\theta}{8\sqrt{2S}}\bk^2 = \frac{3\sqrt{\rho(1-\rho)}}{4\sqrt{2S}}\bk^2
}
for the triangular lattice. Finally, the quartic terms become 
\eqn{
\hH^{(4)} = \frac{1}{N}\sum_{\bk, \bq, \bq'} g(\bk; \bq, \bq')\,\had_{\frac{\bk+\bq'}{2}}\had_{\frac{\bk-\bq'}{2}}\ha_{\frac{\bk-\bq}{2}}\ha_{\frac{\bk+\bq}{2}},
}
where the vertex function is given by
\eqa{
g(\bk; \bq, \bq') = &\frac{1 - \cos\theta}{4S}\frac{1}{N}\sum_{i,j,l} \frac{g_{ijl}}{2} \frac{1}{4}\Biggl[ \cos\left(\frac{\bk+\bq}{2}\cdot(\br_j - \br_l)\right)
+ \cos\left(\frac{\bk-\bq}{2}\cdot(\br_j - \br_l)\right)
+ \cos\left(\frac{\bk+\bq'}{2}\cdot(\br_j - \br_l)\right) \nn \\
&\;\;\;\;\;\;\;\;\;\;\;\;\;+ \cos\left(\frac{\bk-\bq'}{2}\cdot(\br_j - \br_l)\right)
- 2\cos\left(\frac{\bq+\bq'}{2}\cdot(\br_j - \br_l)\right)
- 2\cos\left(\frac{\bq-\bq'}{2}\cdot(\br_j - \br_l)\right) \Biggr] \nn \\
&+ \frac{\cos\theta}{4S}\frac{1}{N}\sum_{i,j,l} \frac{g_{ijl}}{2} \frac{1}{2}\Biggl[
\cos\left(\frac{\bq+\bq'}{2}\cdot(\br_i - \br_j)\right)
+ \cos\left(\frac{\bq-\bq'}{2}\cdot(\br_i - \br_j)\right) \nn \\
&\;\;\;\;\;\;\;\;\;\;\;\;\;\;\;\;\;\;\;\;\;\;\;\;\;\;+ \cos\left(\frac{\bq+\bq'}{2}\cdot(\br_i - \br_l)\right)
+ \cos\left(\frac{\bq-\bq'}{2}\cdot(\br_i - \br_l)\right) \nn \\
&\;\;\;\;\;\;\;\;\;\;\;\;\;-2\cos\left(\frac{\bk}{2}\cdot(\br_j - \br_l)\right)\Biggl[
\cos\left(\frac{\bq}{2}\cdot(\br_i - \br_j)\right)\cos\left(\frac{\bq'}{2}\cdot(\br_i - \br_l)\right) \nn \\
&\;\;\;\;\;\;\;\;\;\;\;\;\;\;\;\;\;\;\;\;\;\;\;\;\;\;\;\;\;\;\;\;\;\;\;\;\;\;\;\;\;\;\;\;\;\;\;\;\;\;\;+ \cos\left(\frac{\bq'}{2}\cdot(\br_i - \br_j)\right)\cos\left(\frac{\bq}{2}\cdot(\br_i - \br_l)\right)\Biggr]
\Biggr].
}
We can again expand this at small momentum, and recover
\eqa{
g(\bk; \bq, \bq') = \frac{1}{N}\sum_{i,j,l} \frac{g_{ijl}}{2}\Biggl[\frac{1 - \cos\theta}{4S}\left(-\frac{1}{8}\left(\bk\cdot(\br_j - \br_l)\right)^2
+ \frac{1}{16}\left(\bq\cdot(\br_j - \br_l)\right)^2
+ \frac{1}{16}\left(\bq'\cdot(\br_j - \br_l)\right)^2 \right) \nn \\
+ \frac{\cos\theta}{4S}\frac{1}{4}\left(\bk\cdot(\br_j - \br_l)\right)^2\Biggr].
}
We thus find
\eqn{
g(k; q, q') \sim -\frac{3(2 - 3\rho)}{8S}k^2 + \frac{3(1 - \rho)}{16S}(q^2 + q'\,\!^2)
}
in the triangular chain, and
\eqn{
g(\bk; \bq, \bq') \sim -\frac{3(2 - 3\rho)}{16S}\bk^2 + \frac{3(1 - \rho)}{32S}(\bq^2 + \bq'\,\!^2)
}
in the triangular lattice.

\section{Asymptotic form of the autocorrelation function}
The numerical results we present extract the diffusion constants from the autocorrelator of the particle density in the classical models,
\eqn{
\mA(t) := \frac{1}{N}\sum_j \avg{(\rho_j(t) - \rho)(\rho_j(0) - \rho)},
}
where $\rho_j(t)$ is the occupancy of site $j$ at time $t$. For completeness, we derive here its general asymptotic form, assuming that the model is diffusive. The diffusion constant $D$ is defined such that the decay rate of the slowest mode is $D\bk^2$. That is, by assumption, 
\eqn{
\hH = \sum_{\bk} D\bk^2 \had_{\bk}\ha_{\bk} + ...\;.
}
Now, writing $\delta\rho_i(t) = \rho_i(t) - \rho$, we can evaluate the autocorrelator with the classical-to-quantum mapping (see for example \cite{jack2006mappings}), which gives
\eqa{
\mA(t) &:= \frac{1}{N}\sum_i\avg{\delta\rho_i(t)\!\;\delta\rho_i(0)} 
= \frac{1}{N}\sum_i\bra{\Psi(\rho)}\delta\hat{\rho}_i\!\;e^{-Ht}\!\;\delta\hat{\rho}_i\ket{\Psi(\rho)}  \nn \\
&= \frac{1}{N}\sum_i\sum_{\{\boldsymbol{\eta}\}} \bra{0}\delta\hat{\rho}_i\!\;e^{-Ht}\Bigl({\had}_1{^{\eta_1}}\dots{\had}_N{^{\eta_N}}\ket{0}\Bigr)\Bigl(\bra{0}{\ha_1}^{\eta_1}\dots{\ha_N}^{\eta_N}\Bigr)\delta\hat{\rho}_i\ket{0}.
} 
In the last line we have inserted a resolution of the identity and written $\ket{\Psi(\rho)} = \ket{0}$. We have also implicitly assumed $t > 0$ and so dropped the explicit imaginary time-ordering symbol.
Now, the operator $\delta \hat{\rho}_i$ is given in terms of the bosons by
\eqa{
\delta\hat{\rho}_i &= S + \hS^z_i - \rho \nn \\
&= S - \rho + S(2\rho - 1) - (2\rho - 1)\had_i\ha_i - \sqrt{\rho(1 - \rho)}\left(\had_i\sqrt{2S - \had_i\ha_i} + \sqrt{2S - \had_i\ha_i}\ha_i\right) \nn \\
&\underbrace{=}_{S = 1/2} - (2\rho - 1)\had_i\ha_i - \sqrt{\rho(1 - \rho)}\left(\had_i\sqrt{1 - \had_i\ha_i} + \sqrt{1 - \had_i\ha_i}\ha_i\right),
\label{eq:delta_rho}
}
and in the last line we have remembered that $S = 1/2$ when mapping from the classical models. Now, consider the matrix elements $\bigl(\bra{0}{\ha_1}^{\eta_1}\dots{\ha_N}^{\eta_N}\bigr)\delta \hat{\rho}_i\ket{0}$. It is clear from Eq.~\eqref{eq:delta_rho} that this matrix element is non-zero only if exactly one of the $\eta_j = 1$, and all the others are zero. And when acting on the vacuum, the square root terms simply return unity. We thus obtain
\eqa{
\mA(t) &= \underbrace{\rho(1 - \rho)}_{=\,\chi}\frac{1}{N}\sum_{i,j} \bra{0}\ha_i e^{-Ht}\had_j\ket{0}\underbrace{\bra{0}\ha_j\had_i\ket{0}}_{=\,\delta_{ij}} \nn \\
&= \frac{\chi}{N}\sum_{i}\bra{0}\ha_i e^{-Ht}\had_i\ket{0} \nn \\
&= \chi \int \frac{d^d\bk}{\Omega_{\mathrm{BZ}}}\frac{d^d\bk'}{\Omega_{\mathrm{BZ}}}\frac{1}{N}\sum_{i} e^{i(\bk - \bk')\cdot\br_i}\bra{0}\ha_{\bk} e^{-Ht}\had_{\bk'}\ket{0} \nn \\
&\sim \chi \int \frac{d^d\bk}{\Omega_{\mathrm{BZ}}}\frac{d^d\bk'}{\Omega_{\mathrm{BZ}}}\frac{1}{N}\sum_{i} e^{i(\bk - \bk')\cdot\br_i}e^{-D\bk'^2 t}\underbrace{\bra{0}\ha_{\bk}\had_{\bk'}\ket{0}}_{=\,\Omega_{\mathrm{BZ}}\,\delta^{(d)}(\bk - \bk')} \nn \\
&= \chi \int \frac{d^d\bk}{\Omega_{\mathrm{BZ}}}e^{-D\bk^2 t}.
\label{eq:autocorr_from_Greens_function}
}
At late times $t \gg 1/D\Lambda^2$, where $\Lambda$ is the ultraviolet scale, we can simply take the integral over $\bk \in \mathbb{R}^d$, which yields
\eqn{
\mA(t) = \frac{(2\pi)^d}{\Omega_{\mathrm{BZ}}}\frac{\chi}{(4\pi D t)^{d/2}} + ...\,.
}

\section{Analytic structure of the Green's function}

In the main text, we showed that (for parity-symmetric, three-site gate models) the diffusion constant obtained from the non-interacting spin wave theory is not corrected; there are no terms in the self-energy at $\mO(k^2)$, and so the diffusion pole is not shifted at this order. 

However, the alert reader may have noticed that the factor $\mF(\omega, k)$ in Eq.~(13) introduces a branch cut along the negative imaginary axis below $-iDk^2/2$. This branch cut is closer to the origin than the diffusion pole, and so might be expected to provide a slower decay rate, effectively reducing the diffusion constant. We address this concern here, and show that the diffusion constant is still determined by the position of the pole. We will also explicitly evaluate the self-energy diagram in Eq.~\eqref{eq:Green's_function_sup}.

Now, the imaginary time-ordered Green's function of a magnon is, in the non-interacting theory,
\eqn{
\mG^{(0)}(t,k) = \bra{\Psi(\rho)}\mathcal{T}_t\,\ha_{k}(t)\had_k(0)\ket{\Psi(\rho)} = \Theta(t)e^{-D(\rho)k^2t},
}
which, after Fourier transform, becomes
\eqn{
\mG^{(0)}(\omega,k)=\int_{-\infty}^\infty dt\, e^{i\omega t}\mG^{(0)}(t,k)=\frac{-1}{i\omega-Dk^2}.
}
Cf. Eq.~\eqref{eq:autocorr_from_Greens_function}, the classical autocorrelator, from which the diffusion constant is measured, can be obtained from the imaginary time-ordered Green's function (at $t > 0$); and, recall, that $t$ denotes real time in the classical model and imaginary time in the corresponding quantum theory.

Note that with this choice of sign convention the Feynman rules do not require any extra factors of $i$ or $-1$ to be attached to vertices or propagators when evaluating the diagrams. The full propagator has the form given by the Dyson equation,
\eqn{
\mG(\omega,k) = \frac{-1}{i\omega-Dk^2+\Sigma(\omega, k)},
}
which we will transform back to the temporal domain to compare with the decay rates of the non-interacting theory,
\eqn{
\mG(t, k) = \int_{-\infty}^{\infty} \frac{d\omega}{2\pi}\,e^{-i\omega t}\mG(\omega, k).
}
Now, as discussed in the main text, the interacting Green's function is given implicitly by
\eqa{
\begin{gathered}
\tik{
\draw[backarrow, double] (0, 0) -- (1.2, 0);
}
\end{gathered}
\;\;=\;\;
\begin{gathered}
\tik{
\draw[backarrow, thick] (0, 0) -- (1.2, 0);
}
\end{gathered}
\;\;+\;\;
\begin{gathered}
\tik{
\draw[forwardarrow, thick] (0.0, 0) -- (-0.7, 0);
\draw[forwardarrow, double] (1.7, 0) -- (1.18, 0);
\draw[backarrow, thick] (0,0) arc (180:0:0.5);
\draw[backarrow, thick] (0,0) arc (-180:0:0.5);
\draw[fill=gray] (0.95, 0.24) -- (1.2, 0.0) -- (0.95, -0.24) -- cycle;
\fill[black] (0,0) circle [radius=2pt];
}
\end{gathered},
\label{eq:Green's_function_sup}
}
where, in the effective field theory, all of the scattering processes that contribute to the self-energy are absorbed into the renormalised $1\to2$ vertex. That is, considering any perturbative diagram (i.e., composed of bare propagators and bare vertices) that contributes to the self-energy, we must eventually end up with two bare propagators recombining into one at the left-hand bare $2\to1$ vertex, and we consider everything else to be a correction to the right-hand $1\to2$ vertex, e.g.,
\eqn{
\dots\;,\;\;
\begin{gathered}
\tik{
\draw[thick] (-0.0, 0) -- (-0.7, 0);
\draw[thick] (2.0, 0) -- (2.7, 0);
\draw[backarrow, thick] (0,0) arc (180:0:0.5);
\draw[backarrow, thick] (0,0) arc (180:360:0.5);
\draw[forwardarrow, thick] (2,0) arc (360:180:0.5);
\draw[forwardarrow, thick] (2,0) arc (360:540:0.5);
\fill[black] (0,0) circle [radius=2pt];
\fill[black] (1,0) circle [radius=2pt];
\fill[black] (2,0) circle [radius=2pt];
\draw[dashed, red] (0.3, 1.2) -- (0.3, -1.2);
\draw[dashed, red] (0.3, 1.2) arc (90:26:2.2);
\draw[dashed, red] (0.3, -1.2) arc (-90:-26:2.2);
}
\end{gathered}
\;,\;\;
\begin{gathered}
\tik{
\draw[thick] (-0.2, 0) -- (-0.7, 0);
\draw[forwardarrow, thick] (0.7, 0.5) -- (0.35, 0.25);
\draw[forwardarrow, thick] (0.35, 0.25) -- (-0.2, 0);
\draw[forwardarrow, thick] (1.0, -1.0) -- (-0.2, 0);
\draw[thick] (2.0, 0) -- (2.7, 0);
\draw[backarrow, thick] (1.65, 0.25) -- (2.0, 0);
\draw[backarrow, thick] (1.3, 0.5) -- (1.65, 0.25);
\draw[backarrow, thick] (1.0, -1.0) -- (1.65, 0.25);
\draw[forwardarrow, thick] (1.0, -1.0) -- (0.35, 0.25);
\draw[backarrow, thick] (1.0, -1.0) -- (2.0, 0);
\draw[backarrow, thick] (0.7,0.5) arc (180:0:0.3);
\draw[backarrow, thick] (0.7,0.5) arc (180:360:0.3);
\fill[black] (-0.2,0) circle [radius=2pt];
\fill[black] (0.7,0.5) circle [radius=2pt];
\fill[black] (0.35,0.25) circle [radius=2pt];
\fill[black] (1.3,0.5) circle [radius=2pt];
\fill[black] (1.0,-1.0) circle [radius=2pt];
\fill[black] (1.65,0.25) circle [radius=2pt];
\fill[black] (2,0) circle [radius=2pt];
\draw[dashed, red] (0.25, 1.2) -- (0.3, -1.2);
\draw[dashed, red] (0.25, 1.2) arc (90:26:2.2);
\draw[dashed, red] (0.25, -1.2) arc (-90:-26:2.2);
}
\end{gathered}
\;,\;\;
\begin{gathered}
\tik{
\draw[thick] (-0.0, 0) -- (-0.7, 0);
\draw[forwardarrow, thick] (0.5, 0.5) -- (0, 0);
\draw[forwardarrow, thick] (0.7, -0.5) -- (0, 0);
\draw[thick] (2.0, 0) -- (2.7, 0);
\draw[backarrow, thick] (1.3, 0.5) -- (2.0, 0);
\draw[backarrow, thick] (1.5, -0.5) -- (2.0, 0);
\draw[backarrow, thick] (0.5,0.5) arc (180:0:0.4);
\draw[backarrow, thick] (0.5,0.5) arc (180:360:0.4);
\draw[forwardarrow, thick] (1.5,-0.5) arc (360:180:0.4);
\draw[forwardarrow, thick] (1.5,-0.5) arc (360:540:0.4);
\fill[black] (0,0) circle [radius=2pt];
\fill[black] (0.5,0.5) circle [radius=2pt];
\fill[black] (0.7,-0.5) circle [radius=2pt];
\fill[black] (1.3,0.5) circle [radius=2pt];
\fill[black] (1.5,-0.5) circle [radius=2pt];
\fill[black] (2,0) circle [radius=2pt];
\draw[dashed, red] (0.3, 1.2) -- (0.3, -1.2);
\draw[dashed, red] (0.3, 1.2) arc (90:26:2.2);
\draw[dashed, red] (0.3, -1.2) arc (-90:-26:2.2);
}
\end{gathered}
\;,\;\;\dots
\;\;\in\;\;
\begin{gathered}
\tik{
\draw[thick] (-0.0, 0) -- (-0.3, 0);
\draw[thick] (1.4, 0) -- (1.18, 0);
\draw[backarrow, thick] (0,0) arc (180:0:0.5);
\draw[backarrow, thick] (0,0) arc (-180:0:0.5);
\draw[fill=gray] (0.95, 0.24) -- (1.2, 0.0) -- (0.95, -0.24) -- cycle;
\fill[black] (0,0) circle [radius=2pt];
}
\end{gathered}\;.
}
Within the effective field theory then, the self-energy is
\eqa{
\Sigma(\omega, k) 
\;=\;
\begin{gathered}
\tik{
\draw[thick] (-0.0, 0) -- (-0.3, 0);
\draw[thick] (1.4, 0) -- (1.18, 0);
\draw[backarrow, thick] (0,0) arc (180:0:0.5);
\draw[backarrow, thick] (0,0) arc (-180:0:0.5);
\draw[fill=gray] (0.95, 0.24) -- (1.2, 0.0) -- (0.95, -0.24) -- cycle;
\fill[black] (0,0) circle [radius=2pt];
\draw (0.5,0.5) node[above]{$\frac{k+q}{2}$};
\draw (0.5,-0.5) node[below]{$\frac{k-q}{2}$};
\draw (1.4,0) node[right]{$k$};
\draw (-0.3,0) node[left]{$k$};
}
\end{gathered}
\;&=\;
\int \frac{d\omega_q}{2\pi}\frac{dq}{2\pi} (\lambda_{(3)}k^2)(\tilde{\lambda}_{(3)}k^2)
\frac{-1}{\frac{i\omega+i\omega_q}{2} - D\left(\frac{k+q}{2}\right)^2}
\frac{-1}{\frac{i\omega-i\omega_q}{2} - D\left(\frac{k-q}{2}\right)^2} \nn \\
\;&=\;
\int \frac{dq}{2\pi} \frac{4\lambda_{(3)}\tilde{\lambda}_{(3)} k^4}{Dk^2 + Dq^2 - 2i\omega}
\;=\;
\frac{2\lambda_{(3)}\tilde{\lambda}_{(3)}}{\sqrt{D}}\frac{k^4}{\sqrt{Dk^2 - 2i\omega}},
\label{eq:self_energy_renormalised_vertex}
}
and the full one-magnon Green's function is
\eqn{
\mG(\omega,k) = \frac{-1}{i\omega-Dk^2+ \frac{2\lambda_{(3)}\tilde{\lambda}_{(3)}}{\sqrt{D}}\frac{k^4}{\sqrt{Dk^2 - 2i\omega}} + \dots}.
}

\begin{figure}[t]
\centering
\subfigure[Original contour]{\begin{tikzpicture}[>=latex]
    \draw[->] (-3,0) -- (3,0) node[right] {$\Re \omega$};
    \draw[->] (0,-3) -- (0,3) node[above] {$\Im \omega$};

    \draw[very thick, red, dashed] (0,-3) -- (0,-1);
    \fill[red] (0,-1) circle (2pt) node[above left] {$-iDk^2/2$};

    \fill (0,0) circle (2pt) node[below left] {0};

    \draw[dashed] (1,-2.5) circle (4pt);
    \node at (1.7,-2.1) {$-iDk^2+\frac{2\lambda_{(3)}\tilde\lambda_{(3)}}{D}k^3$};
    \draw[dashed] (-1,-2.5) circle (4pt);
    \node at (-1.7,-2.1) {$-iDk^2-\frac{2\lambda_{(3)}\tilde\lambda_{(3)}}{D}k^3$};

    \draw[blue, thick,
        postaction={decorate},
        decoration={
          markings,
          mark=at position 0.25 with {\arrow{>}},
          mark=at position 0.5 with {\arrow{>}},
          mark=at position 0.75 with {\arrow{>}}
        }
    ] (-3,0.2) -- (3,0.2);
\end{tikzpicture}
}
\hspace{1cm}
\subfigure[Deformed contour]{\begin{tikzpicture}[>=latex]
    \draw[->] (-3,0) -- (3,0) node[right] {$\Re \omega$};
    \draw[->] (0,-3) -- (0,3) node[above] {$\Im \omega$};

    \draw[very thick, red, dashed] (0,-3) -- (0,-1);
    \fill[red] (0,-1) circle (2pt) node[above left] {$-iDk^2/2$};

    \fill (0,0) circle (2pt) node[below left] {0};

    \draw[thick, dashed, blue] (1,-2.5) circle (4pt);
    \node at (1.7,-2.1) {$-iDk^2+\frac{2\lambda_{(3)}\tilde\lambda_{(3)}}{D}k^3$};
    \draw[dashed] (-1,-2.5) circle (4pt);
    \node at (-1.7,-2.1) {$-iDk^2-\frac{2\lambda_{(3)}\tilde\lambda_{(3)}}{D}k^3$};

    \draw[blue, thick,
        postaction={decorate},
        decoration={
          markings,
          mark=at position 0.25 with {\arrow{>}},
          mark=at position 0.5 with {\arrow{>}},
          mark=at position 0.75 with {\arrow{>}}
        }
    ] (0,-0.8) -- (3,-0.8);

    \draw[blue, thick, dashed,
        postaction={decorate},
        decoration={
          markings,
          mark=at position 0.25 with {\arrow{<}},
          mark=at position 0.5 with {\arrow{<}},
          mark=at position 0.75 with {\arrow{<}}
        }
    ] (0,-1.2) -- (3,-1.2);

    \draw[blue, thick] (0,-0.8) arc[start angle=90, end angle=270, radius=0.2];
\end{tikzpicture}
}

\caption{The original contour of integration $\mathcal{C}$ and the deformed one. There are two poles with opposite real part on the second Riemann sheet. Deforming the contour picks the pole with positive real part and the cut contribution is the difference between the values of the integrand on the upper and lower plane.}
\label{fig:analytic_G}
\end{figure}
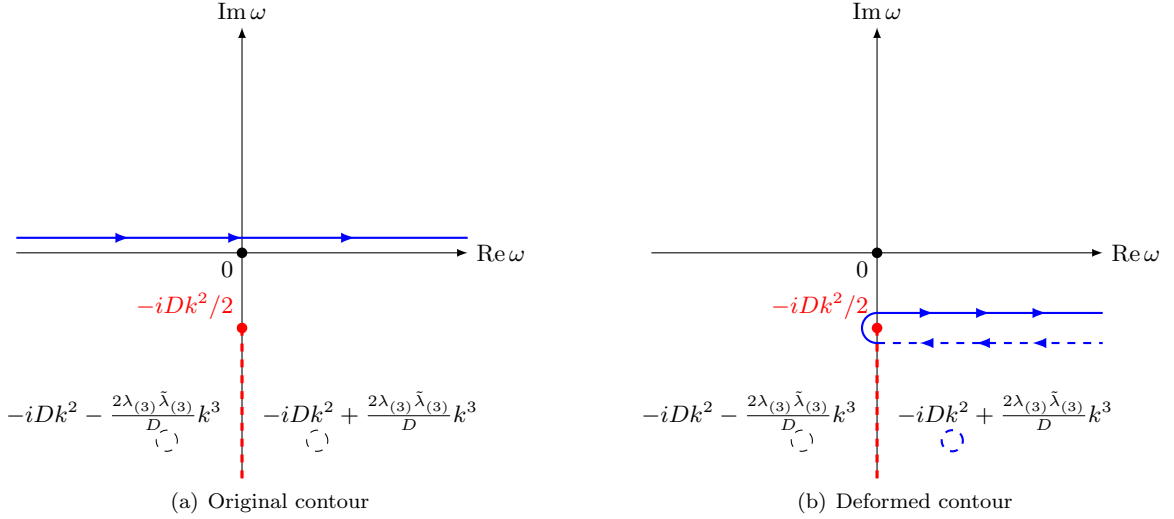

The propagator in the non-interacting theory has a single simple pole in the lower half-plane at $\omega_* = -iDk^2$; the analytic structure in the interacting theory is more involved, and sheds light on several of its features. 
There is a branch cut along the negative imaginary axis below $-iDk^2/2$, and there are no other singularities on the first Riemann sheet; the diffusion pole, which originally lay on this cut, splits into two poles which appear with opposite real part on the second Riemann sheet, at $\omega_*^{\pm} = -iDk^2 \pm Ak^3 + ...\,$, where $A = 2\lambda_{(3)}\tilde\lambda_{(3)}/\sqrt{D}$ (see, e.g., \cite{PhysRevLett.122.091602}).   

When we now transform $\mG(\omega, k)$ back to the temporal domain, we will indeed pick up a contribution from the (shifted) pole $\omega^+_*$, but also from the branch cut. We show in Fig.~\ref{fig:analytic_G} how the integration contour can be deformed to see this.

Let us now explicitly write out the two contributions,
\eqn{
\mG(t, k) = \int_{-\infty}^{\infty}\frac{d\omega}{2\pi}\frac{-e^{-i\omega t}}{i\omega-Dk^2+ \frac{2\lambda_{(3)}\tilde{\lambda}_{(3)}}{\sqrt{D}}\frac{k^4}{\sqrt{Dk^2 - 2i\omega}} + \dots} 
= \mZ e^{-i\omega_*^+ t} + \int_{\frac{-iDk^2}{2}}^{\frac{-iDk^2}{2} + \infty}\frac{d\omega}{2\pi}e^{-i\omega t}\mB(\omega),
}
where
\eqn{
\mZ = \frac{1}{1 - \frac{2i\lambda_{(3)}\tilde\lambda_{(3)}|k|}{D^2} +\mO(k^2)}
}
is the residue at the pole $\omega_*^+$ on the second Riemann sheet, and
\eqn{
\mB(\omega) = \frac{\frac{4\lambda_{(3)}\tilde\lambda_{(3)}}{\sqrt{D}} k^4\sqrt{Dk^2 - 2i\omega}}{-\frac{4\lambda_{(3)}^2\tilde\lambda_{(3)}^2}{D} k^8 + \left(Dk^2 - i\omega\right)^2\left(Dk^2 - 2i\omega\right)}
}
is the difference between $\mG(\omega, k)$ on the first Riemann sheet and the second (where $\sqrt{\dots} \,\mapsto -\sqrt{\dots}$).

The true decay of a mode $k$ is no longer the simple exponential of the non-interacting case; it is the sum of two terms which approximately decay like exponentials,
\eqn{
\mG(t, k)= \mG_{\mathrm{pole}}(t, k) + \mG_{\mathrm{cut}}(t, k).
}
The pole contribution is, asymptotically (at small $k$), the same as in the non-interacting case,
\eqn{
\mG_{\mathrm{pole}}(t, k) = \mZ e^{-i\omega_*^+ t} = \mZ e^{-Dk^2 t} \sim e^{-Dk^2 t}.
}
On the other hand, the contribution from the branch cut is
\eqn{
\mG_{\mathrm{cut}}(t, k) = e^{-Dk^2 t/2} \int_0^{\infty} \frac{dr}{2\pi} e^{-irt} \mB\left(\frac{-iDk^2}{2} + r\right).
}

Now, at large $t$ the factor $e^{-irt}$ will be rapidly oscillating, and kill all contributions away from $r \approx 0$. Expanding $\mB$ near $r = 0$, then, we have
\eqn{
\mB\left(\frac{-iDk^2}{2} + r\right) = -\frac{(1 - i)\sqrt{D}}{\lambda_{(3)}\tilde\lambda_{(3)} k^4}\sqrt{r} \,+\, \mO(r^{3/2}),
}
and so 
\eqn{
\mG_{\mathrm{cut}}(t \gg \tau_\lambda, k) \sim e^{-Dk^2 t/2} \int_0^{\infty} \frac{dr}{2\pi} e^{-irt - 0^+ r}\frac{-(1 - i)\sqrt{D}}{\lambda_{(3)}\tilde\lambda_{(3)} k^4}\sqrt{r} = \left(\frac{\tau_{\lambda}}{t}\right)^{3/2} e^{-Dk^2 t/2}, \;\;\; \tau_{\lambda} = \frac{\pi^{1/3}D^{1/3}}{2^{1/3}\lambda_{(3)}^{2/3}\tilde\lambda_{(3)}^{2/3} k^{8/3}},
}
where $\tau_{\lambda}$ is an interaction timescale. Notice that due to the $k^{-8/3}$ dependence, in the $k\to 0$ limit we have $\tau_{\lambda}\gg \tau_{D}$, where $\tau_D = 1/Dk^2$ is the Thouless time. This result is valid for $t\gg \tau_\lambda$; at short times $t \ll \tau_\lambda$ the prefactor is instead $\sim k^4$ (no $t$ dependence). The crossover between the two types of decay therefore takes place at a time $t^*$ such that
${k^4 e^{-t^*/2\tau_D} \sim e^{-t^*/\tau_D}}$. This gives us 
\eqn{
t^* \sim \tau_D \log\left(\frac{1}{k^2}\right).
}
That is, in the $k\to0$ limit, the time $t^*$ after which we would observe the $e^{-Dk^2 t/2}$ decay from the branch cut instead of the $e^{-Dk^2 t}$ decay from the pole diverges parametrically faster than the Thouless time $\tau_D$. The contribution from the branch cut can thus never be observed for the longest wavelength modes $k \sim 1/L$, and the decay is dominated by the pole contribution.

There is thus no change to the observed diffusion constant.
\vspace{0.2cm}

Let us briefly address the case of higher dimensions. The procedure is the same, but in $d > 1$ the self-energy diagram \eqref{eq:self_energy_renormalised_vertex} is ultraviolet divergent. We will use a soft regularisation scheme that does not break the (emergent) $SO(d)$ symmetry, writing
\eqn{
\frac{1}{X} \to \frac{e^{-X/\Omega}}{X} = \int_{1/\Omega}^{\infty} ds\,e^{-sX},
\label{eq:regularisation_scheme}
}
where $\Omega = D\Lambda^2$ is a frequency cutoff (and the factors $X$ are $\mO(k^2)$, cf. the bare propagator). We thus have
\eqa{
\begin{gathered}
\tik{
\draw[thick] (-0.0, 0) -- (-0.3, 0);
\draw[thick] (1.4, 0) -- (1.18, 0);
\draw[backarrow, thick] (0,0) arc (180:0:0.5);
\draw[backarrow, thick] (0,0) arc (-180:0:0.5);
\draw[fill=gray] (0.95, 0.24) -- (1.2, 0.0) -- (0.95, -0.24) -- cycle;
\fill[black] (0,0) circle [radius=2pt];
\draw (0.5,0.5) node[above]{$\frac{\bk+\bq}{2}$};
\draw (0.5,-0.5) node[below]{$\frac{\bk-\bq}{2}$};
\draw (1.4,0) node[right]{$\bk$};
\draw (-0.3,0) node[left]{$\bk$};
}
\end{gathered}
\;=\;
\int \frac{d^d\bq}{(2\pi)^d} \frac{4\lambda_{(3)}\tilde{\lambda}_{(3)} \bk^4}{D\bk^2 + D\bq^2 - 2i\omega}
\;\sim\;
\int_{1/\Omega}^{\infty} ds \int \frac{d^d\bq}{(2\pi)^d} \,4\lambda_{(3)}\tilde{\lambda}_{(3)} \bk^4 e^{-s(D\bk^2 + D\bq^2 - 2i\omega)} \nn \\
\;=\;
4\lambda_{(3)}\tilde{\lambda}_{(3)} \bk^4 \int_{1/\Omega}^{\infty} ds \,e^{-s(D\bk^2 - 2i\omega)} (4\pi D s)^{-d/2}
\;=\; 
4\lambda_{(3)}\tilde{\lambda}_{(3)} \bk^4 (4\pi D)^{-d/2} \alpha^{d/2 - 1}\,\Gamma\left(1 - \frac{d}{2}, \frac{\alpha}{\Omega}\right),
}
where $\alpha := D\bk^2 - 2i\omega$, and $\Gamma$ is the incomplete Gamma function. One could, if so inclined, convert this to a dimensional regularisation by taking $\Omega \to \infty$ but interpreting $d$ as a continuous real number. If we wish to take integer $d$, however, we need to keep the cutoff and look at the leading $\Omega$ divergence. For the case $d = 2$, this leading divergence is $\log(\Omega/\alpha)$, so the leading term in the self-energy is $\sim \bk^4 \log(\Omega/(D\bk^2 - 2i\omega))$. The same considerations as in the one-dimensional case show that the diffusion constant is similarly not corrected.

\section{Vertex Corrections}

In the main text, and the previous section of this supplementary, we argued that within the effective field theory the renormalised $1\to2$ vertex retains the same form at low momentum, $\lambda_{(3)}k^2 \to \tilde\lambda_{(3)}k^2$. From this we obtain the lowest order (in $k$) contribution to the self-energy, and, up to the same order, the (form of the) exact Green's function.

To give some further confidence that the vertex renormalises in the way we expect, we compute here the perturbative contributions to the self-energy up to $\mO(1/S^2)$, and observe they are consistent with the inferred form of the vertex in Eq.~\eqref{eq:Green's_function_sup}. Recall that $\lambda_{(n)},\gamma_{(n)} \in \mO(S^{-(n-2)/2})$.

The $\mO(1/S)$ term is just the diagram we have already computed, but with two bare vertices:
\eqn{
\begin{gathered}
\tik{
\draw[backarrow, thick] (0,0) arc (180:0:0.5);
\draw[backarrow, thick] (0,0) arc (-180:0:0.5);
\draw[thick] (0, 0) -- (-0.2, 0);
\draw[thick] (1.2, 0) -- (1.0, 0);
\fill[black] (0,0) circle [radius=2pt];
\fill[black] (1,0) circle [radius=2pt];
\draw (0.5,0.5) node[above]{$\frac{k+q}{2}$};
\draw (0.5,-0.5) node[below]{$\frac{k-q}{2}$};
\draw (1.2,0) node[right]{$k$};
\draw (-0.2,0) node[left]{$k$};
}
\end{gathered}
\;=\;
\frac{2\lambda_{(3)}^2}{\sqrt{D}}\frac{k^4}{\sqrt{Dk^2 - 2i\omega}} + ...\,
\;\in\;
\mO(1/S).
\label{eq:self_energy_first_order}
}

At $\mO(1/S^2)$ there are three diagrams. The first of these is just two-particle scattering,
\eqa{
\begin{gathered}
\tik{
\draw[backarrow, thick] (0,0) arc (180:0:0.5);
\draw[backarrow, thick] (0,0) arc (-180:0:0.5);
\draw[backarrow, thick] (1,0) arc (180:0:0.5);
\draw[backarrow, thick] (1,0) arc (-180:0:0.5);
\draw[thick] (0, 0) -- (-0.2, 0);
\draw[thick] (2.2, 0) -- (2.0, 0);
\fill[black] (0,0) circle [radius=2pt];
\fill[black] (1,0) circle [radius=2pt];
\fill[black] (2,0) circle [radius=2pt];
\draw (1.5,0.5) node[above]{$\frac{k+q}{2}$};
\draw (1.5,-0.5) node[below]{$\frac{k-q}{2}$};
\draw (0.5,0.5) node[above]{$\frac{k+q'}{2}$};
\draw (0.5,-0.5) node[below]{$\frac{k-q'}{2}$};
\draw (2.2,0) node[right]{$k$};
\draw (-0.2,0) node[left]{$k$};
}
\end{gathered}
\;&=\;
\int \frac{d\omega_q}{2\pi}\frac{d\omega_q'}{2\pi}\frac{dq}{2\pi}\frac{dq'}{2\pi} \left[(\lambda_{(3)}k^2)^2 (\lambda_{(4)}k^2 + \gamma_{(4)}q^2 + \gamma_{(4)}q'^2)
\frac{-1}{\frac{i\omega+i\omega_q}{2} - D\left(\frac{k+q}{2}\right)^2}\right. \nn \\[-0.4cm]
&\;\;\;\;\;\;\;\;\;\;\;\;\;\;\;\;\;\;\;\;\;\;\;\;\;\;\;\;\left.\times\; \frac{-1}{\frac{i\omega-i\omega_q}{2} - D\left(\frac{k-q}{2}\right)^2}
\frac{-1}{\frac{i\omega+i\omega_q'}{2} - D\left(\frac{k+q'}{2}\right)^2}
\frac{-1}{\frac{i\omega-i\omega_q'}{2} - D\left(\frac{k-q'}{2}\right)^2}\right] \nn \\[0.2cm]
\;&=\;
\int \frac{dq}{2\pi}\frac{dq'}{2\pi} \frac{16\lambda_{(3)}^2 k^4 (\lambda_{(4)}k^2 + \gamma_{(4)}q^2 + \gamma_{(4)}q'^2)}{(Dk^2 + Dq^2 - 2i\omega)(Dk^2 + Dq'^2 - 2i\omega)} \nn \\[0.2cm]
\;&\sim\;
\int_{1/\Omega}^{\infty} ds ds' \int \frac{dq}{2\pi}\frac{dq'}{2\pi} 16\lambda_{(3)}^2 k^4 (\lambda_{(4)}k^2 + \gamma_{(4)}q^2 + \gamma_{(4)}q'^2) e^{-s(Dk^2 + Dq^2 - 2i\omega)}e^{-s'(Dk^2 + Dq'^2 - 2i\omega)} \nn \\[0.2cm]
\;&=\;
\frac{8\lambda_{(3)}^2\gamma_{(4)}\sqrt{\Omega}}{\sqrt{\pi}D^2} \frac{k^4}{\sqrt{Dk^2 - 2i\omega}} + \mO(k^4)
\;\in\;\mO(1/S^2),
\label{eq:self_energy_second_order_diagram_one}
}
where we have used the soft regularisation scheme \eqref{eq:regularisation_scheme}. This is exactly what we would obtain from Eq.~\eqref{eq:self_energy_renormalised_vertex} with 
\eqn{
\tilde\lambda_{(3)} = \lambda_{(3)} + \frac{4\lambda_{(3)}\gamma_{(4)}\Lambda}{\sqrt{\pi}D} + \mO(S^{-5/2}),
}
where we have used $\Omega = D\Lambda^2$. 

The second diagram is
\eqa{
\begin{gathered}
\tik{
\draw[thick] (0, 0) -- (-0.7, 0);
\draw[thick] (3, 0) -- (3.7, 0);
\draw[backarrow, thick] (0,0) arc (180:90:1.5);
\draw[backarrow, thick] (1.5,1.5) arc (90:0:1.5);
\draw[backarrow, thick] (0,0) arc (180:270:1.5);
\draw[backarrow, thick] (1.5,-1.5) arc (270:360:1.5);
\draw[forwardarrow, thick] (1.8, -1.5) -- (1.2, 1.5);
\fill[black] (0,0) circle [radius=2pt];
\fill[black] (3,0) circle [radius=2pt];
\fill[black] (1.8,-1.463) circle [radius=2pt];
\fill[black] (1.2,1.463) circle [radius=2pt];
\draw (-0.7, 0) node[left] {$k$};
\draw (3.7, 0) node[right] {$k$};
\draw (0.5, 1.4) node[left] {$\frac{k+q}{2}+q'$};
\draw (0.5, -1.4) node[left] {$\frac{k-q}{2}-q'$};
\draw (2.5, 1.4) node[right] {$\frac{k+q}{2}$};
\draw (2.5, -1.4) node[right] {$\frac{k-q}{2}$};
\draw (1.6, 0) node[right] {$q'$};
}
\end{gathered}
\;=\;
\lambda_{(3)}^4 k^4 \int_{-\Lambda}^{\Lambda} \frac{dq'}{2\pi} \frac{q'\,\!^4}{\sqrt{D(3Dq'\,\!^2 + \alpha^2)}\,(Dq'\,\!^2 + \alpha^2)(2Dq'\,\!^2 + \frac{1}{2}\alpha^2)} + ...\,
\;\in\;
\mO(k^4),
}
where, again, $\alpha := \sqrt{Dk^2 - 2i\omega}$. This diagram is logarithmically divergent in the cutoff $\Lambda$ and gives contributions of order $k^4\log(\Lambda/\alpha)$ (there are also terms $k^4\alpha$, $k^4\alpha^2$, etc., but no terms $\sim k^4/\alpha \in \mO(k^3)$). So this diagram does not change the diffusion coefficient either. 

The third and last diagram at $\mO(1/S^2)$ is
\eqa{
\begin{gathered}
\tik{
\draw[thick] (-0.0, 0) -- (-0.7, 0);
\draw[forwardarrow, thick] (1, 1) -- (0, 0);
\draw[thick] (4, 0) -- (4.7, 0);
\draw[backarrow, thick] (3, 1) -- (4, 0);
\draw[backarrow, thick] (1,1) arc (180:0:1);
\draw[backarrow, thick] (1,1) arc (180:360:1);
\draw[backarrow, thick] (0,0) arc (220:320:2.64);
\fill[black] (0,0) circle [radius=2pt];
\fill[black] (1,1) circle [radius=2pt];
\fill[black] (3,1) circle [radius=2pt];
\fill[black] (4,0) circle [radius=2pt];
\draw (2,2) node[above] {$\frac{k+q}{4} + \frac{q'}{2}$};
\draw (2,0.2) node[above] {$\frac{k+q}{4} - \frac{q'}{2}$};
\draw (2,-1) node[below] {$\frac{k-q}{2}$};
\draw (0.7,0.8) node[left] {$\frac{k+q}{2}$};
\draw (3.3,0.8) node[right] {$\frac{k+q}{2}$};
\draw (-0.7, 0) node[left] {$k$};
\draw (4.7, 0) node[right] {$k$};
}
\end{gathered}
\;=\;
\lambda_{(3)}^4 k^4 \int_{-\Lambda}^{\Lambda} \frac{dq'}{2\pi} \frac{2q'\,\!^4}{\sqrt{D(3Dq'\,\!^2 + 4\alpha^2)}\,(Dq'\,\!^2 + \alpha^2)^2} + ...\,
\;\in\;
\mO(k^4).
}
Analogously to the previous one, this diagram is logarithmically divergent and gives a contribution $\sim k^4\log(\Lambda/\alpha)$ thus not correcting the diffusion coefficient. We thus observe that the assumed low-momentum form of the $1\to2$ vertex in the effective field holds at least up to $\mO(1/S^2)$ in perturbation theory -- based on the general arguments that it must be even in $k$, and terms that generate a mass for the magnon field are forbidden, we conjecture this holds to all orders in perturbation theory.

\section{Diagram generating corrections to the diffusion coefficient}

Let us now explicitly evaluate one of the diagrams that reduces the diffusion constant in the four-site gate chains. Recall that for those models, we have terms in the spin-wave expansion that can cause the direct decay of one magnon into three,
\eqa{
\hH^{(4)'} = \frac{1}{N}\sum_{k,p,q} h(k; q, q')\;\!\had_{\frac{k+q-q'}{3}}\;\!\had_{\frac{k-2q}{3}}\;\!\had_{\frac{k+q+q'}{3}}\;\!\ha_k + \mathrm{h.c.},
}
where the vertex function $h(k; q, q')$ is given explicitly by
\eqa{
h(k; q, q') = &\frac{\rho(1 - \rho)}{4S}\left(\sum_{a, b}\Gamma_{ab}\sigma_a\sigma_b\right) \nn \\
&\times \frac{4}{3}\sin\left(\frac{k}{2}\right) 
\left(\cos\left(\frac{2q'}{3}\right)\sin\left(\frac{k-8q}{6}\right) + 
\cos\left(q'\right)\sin\left(\frac{k-2q}{6}\right) +
\cos\left(\frac{q'}{3}\right)\sin\left(\frac{k+10q}{6}\right)\right).
}
Unlike $f(k; q)$ and $g(k; q, q')$, parity does not constrain this vertex to be even in the external momentum $k$, so terms that are linear in $k$ can appear. At fourth (total) order in momentum, a term
\eqn{
\zeta_{(4)} k\,\!q (q - q') (q + q'), \;\;\; \zeta_{(4)} = -\frac{20}{81}\frac{\rho(1 - \rho)}{4S}\left(\sum_{a, b}\Gamma_{ab}\sigma_a\sigma_b\right),
}
appears in $h(k; q, q')$. This generates an $\mO(k^2)$ term in the self-energy in the two-loop diagram,
\eqa{
\Sigma_{1\to3\to1}(\omega, k)
\;&=\;
\begin{gathered}
\tik{
\draw[backarrow, thick] (0,0) arc (180:0:1.0);
\draw[backarrow, thick] (0,0) arc (-180:0:1.0);
\draw[backarrow, thick] (0, 0) -- (2.0, 0);
\draw[thick] (0, 0) -- (-0.4, 0);
\draw[thick] (2.4, 0) -- (2.0, 0);
\fill[black] (0,0) circle [radius=2pt];
\fill[black] (2.0,0) circle [radius=2pt];
\draw (1.0, 0.0) node[above] {\footnotesize$(k - 2q)/3$};
\draw (1.0, 1.0) node[above] {\footnotesize$(k + q + q')/3$};
\draw (1.0, -1.0) node[below] {\footnotesize$(k + q - q')/3$};
\draw (2.4, 0) node[right] {$k$};
\draw (-0.4, 0) node[left] {$k$};
}
\end{gathered}
\;=\;
\int \frac{d\omega_q}{2\pi}\frac{d\omega_q'}{2\pi}\frac{dq}{2\pi}\frac{dq'}{2\pi} (\zeta_{(4)}k\,\!q(q - q')(q + q'))^2 \left[ \frac{-1}{\frac{i\omega - 2i\omega_q}{3} - D^{(0)}\left(\frac{k - 2q}{3}\right)^2} \right. \nn \\[-0.6cm]
&\;\;\;\;\;\;\;\;\;\;\;\;\;\;\;\;\;\;\;\;\;\;\;\;\;\;\;\;\;\;\;\;\;\;\;\;\;\;\;\;\;\;\;\;\;\;\;\;\;\;\;\;\;\;\;\;\;\;\;\;\;\left. \times\; \frac{-1}{\frac{i\omega + i\omega_q + i\omega_q'}{3} - D^{(0)}\left(\frac{k + q + q'}{3}\right)^2}
\frac{-1}{\frac{i\omega + i\omega_q - i\omega_q'}{3} - D^{(0)}\left(\frac{k + q - q'}{3}\right)^2}\right] \nn \\[0.2cm]
&=\; \int \frac{dq}{2\pi}\frac{dq'}{2\pi}\,\frac{81\zeta_{(4)}^2 k^2 q^2 (q^2 - q'\;\!^2)^2}{2(3D^{(0)}k^2 + 6D^{(0)}q^2 + 2D^{(0)}q'\;\!^2 - 9i\omega)} \nn \\[0.2cm]
&\sim\; \int_{1/\Omega}^{\infty} ds \int \frac{dq}{2\pi}\frac{dq'}{2\pi}\,81\zeta_{(4)}^2 k^2 q^2 (q^2 - q'\;\!^2)^2 e^{-2s(3D^{(0)}k^2 + 6D^{(0)}q^2 + 2D^{(0)}q'\;\!^2 - 9i\omega)} \nn \\[0.2cm]
&=\; \int_{1/\Omega}^{\infty} ds \frac{3\sqrt{3} \zeta_{(4)}^2 k^2 e^{-6s(Dk^2 - 3i\omega)}}{1024 \pi s^4 {D^{(0)}}^4} = \frac{\sqrt{3} \Omega^3 \zeta_{(4)}^2 k^2}{1024\pi {D^{(0)}}^4} + \mO(k^4) \;=\; \underbrace{\frac{\sqrt{3} \Lambda^6 \zeta_{(4)}^2}{1024\pi D^{(0)}}}_{\small \delta D}k^2 + \mO(k^4),
}
where we have again used the soft regularisation, with $\Omega \sim D\Lambda^2$. This diagram thus contributes a finite, negative correction to the diffusion constant, $D^{(0)} \to D^{(0)} - \delta D$.

\section{Subdiffusion in the XOR and XNOR chains}

We close by presenting the numerical evidence that the XOR and XNOR chains (cf. Fig.~3) are subdiffusive at all densities $\rho$ (except $\rho = 0$ and $\rho = 1$ in the XNOR chain), with the scaling relation $x \sim t^{1/4}$. The autocorrelation function has the asymptotic form
\eqn{
\mA(t) \sim \frac{\chi}{(4\pi \lambda t)^{1/4}},
}
for some generalised diffusion coefficient $\lambda$. That the XNOR chain is subdiffusive at $\rho = 1/2$ was shown in Ref.~\cite{singh2021subdiffusion}, and can be explained in terms of an intuitive screening argument. The fate of the transport away from half-filling is less obvious.

\begin{figure*}
    \centering
    \includegraphics[width=\linewidth]{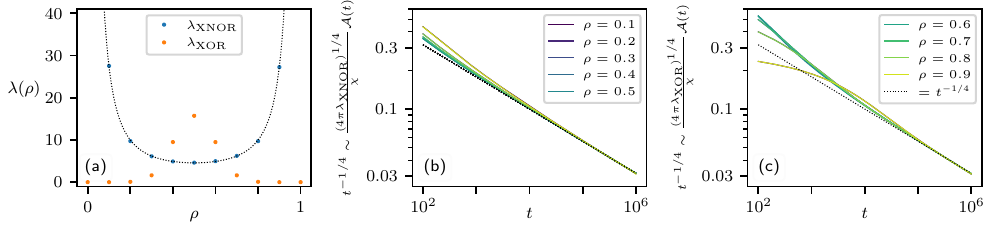}
    \caption{Subdiffusion ($z = 4$) in the XNOR and XOR chains. (a) Generalised diffusion coefficients $\lambda(\rho)$; the dotted line is the divergence of $\lambda_{\mathrm{XNOR}}(\rho)$ as $\rho \to 0, 1$, which fits well numerically to $\lambda_{\mathrm{XNOR}}(\rho) \approx 0.4/(\rho(1 - \rho))^{7/4}$. (b), (c) Autocorrelators of the XNOR and XOR chains, respectively, rescaled to show their consistency with the subdiffusive power law $\mA(t) \sim t^{-1/4}$. $N = 8192$ in all cases.}
    \label{fig:XNOR_and_XOR}
\end{figure*}

We show in Fig.~\ref{fig:XNOR_and_XOR} that at all densities the late time autocorrelators of the XNOR and XOR chains follow a power-law decay $\sim t^{-1/4}$, consistent with subdiffusion. We also show the generalised diffusion coefficients $\lambda$, which have the correct behaviour in the limits $\rho \to 0$ and $\rho \to 1$: $\lambda_{\mathrm{XNOR}} \to \infty$, since the XNOR chain is trivially diffusive in these limits; and $\lambda_{\mathrm{XOR}} \to 0$ as the XOR chain becomes trivially completely frozen.

\end{document}